\documentclass[vecphys]{svmult}

\usepackage{graphicx}        
\usepackage{multicol}        

\makeindex             

\usepackage{amssymb,latexsym,afterpage,calc,array,tabularx,amsmath,eepic,epic,subfigure}

\usepackage[T1]{fontenc}

\usepackage[dvips]{rotating}

\usepackage{cite}

\usepackage{url}

\begin{document}


\newcommand{\mbfr}{\mathbf r}
\newcommand{\mbfc}{\mathbf c}
\newcommand{\mbfd}{\mathbf d}
\newcommand{\mbfe}{\mathbf e}
\newcommand{\mbfv}{\mathbf v}
\newcommand{\mbff}{\mathbf f}
\newcommand{\mbfmu}{\boldsymbol{\mu}}
\newcommand{\mbfxi}{\boldsymbol{\xi}}
\def\vmax{v_{\rm max}}


\title*{Evacuation Dynamics: Empirical Results, Modeling and Applications}

\author{Andreas Schadschneider \inst{1,2} \and Wolfram Klingsch\inst{3}
\and Hubert Kl\"upfel \inst{4} \and Tobias Kretz \inst{5} \and
Christian Rogsch \inst{3} 
\and Armin Seyfried \inst{6}
}

\institute{Institut f\"ur Theoretische Physik, Universit\"at zu K\"oln,
50937 K\"oln,  Germany\\
{\tt as@thp.uni-koeln.de}
\and
Interdisziplin\"ares Zentrum f\"ur komplexe Systeme,
53117 Bonn, Germany
\and
Institute for Building Material Technology and Fire Safety Science, 
University of Wuppertal, 42285 Wuppertal, Germany
\and
TraffGo HT GmbH, Bismarckstr.~142 , 47057 Duisburg, Germany
\and
PTV AG, Stumpfstr.~1, 76131 Karlsruhe, Germany
\and
J\"ulich Supercomputing Centre,
Research Centre J\"ulich, 52425 J\"ulich, Germany
}

\authorrunning{A. Schadschneider et al.}

\maketitle
\vspace{0.5cm}
{\bf to appear in: ``Encyclopedia of Complexity and System Science'',
B.~Meyers (Ed.) (Springer, Berlin, 2008)}

\section{Definition of the subject and its importance}
\label{sec-def}

Today, there are many occasions where a large number of people
gathers in a rather small area. Office buildings and apartment
house become larger and more complex. Very large events related
to sports, entertainment or cultural and religious events
are held all over the world on a regular basis.
This brings about serious safety issues for the participants and
the organizers who have to be prepared for any case of emergency
or critical situation. Usually in such cases the participants
have to be guided away from the dangerous area as fast as possible.
Therefore the understanding of the dynamics of large groups of
people is very important.

In general, evacuation is the egress from an area, a building or
vessel due to a potential or actual threat. In the cases described
above the dynamics of the evacuation processes is quite complex
due to the large number of people and their interaction, external
factors like fire etc., complex building geometries,...
Evacuation dynamics has to be described and understood on different levels:
physical, physiological, psychological, and social. Accordingly, the
scientific investigation of evacuation dynamics involves many research
areas and disciplines.  The system ``evacuation process'' (i.e.\ the
population and the environment) can be modelled on many different
levels of detail, ranging from hydro-dynamic models to artifical
intelligence and multi-agent systems.  There are at least three
aspects of evacuation dynamics that motivate its scientific
investigation: 1) as in most many-particle systems several interesting
collective phenomena can be observed that need to be explained;
2) models need to be developed that are able to reproduce pedestrian 
dynamics in a realistic way, and 3) the application of
pedestrian dynamics to facility design and emergency preparation and
management.

The investigation of evacuation dynamics is a difficult problem that
requires close collaboration between different fields. The origin
of the apparent complexity lies in the fact that one is concerned
with a many-`particle' system with complex interactions that are
not fully understood. Typically the systems are far from equilibrium
and so are e.g.\ rather sensitive to boundary conditions. Motion
and behaviour are influenced by several external factors and often
crowds can be rather inhomogeneous.

In this article we want to deal with these problems from different
perspectives and will not only review the theoretical background,
but also discuss some concrete applications.


\section{Introduction}
\label{sec-intro}

The awareness that emergency exits are one of the most important
factors to ensure the safety of persons in buildings can be traced
more than 100 years. The desasters due to the
fires in the Ringtheater in Vienna and the urban theater in Nizza at
1881 with several hundred fatalities lead to a rethinking of the
safety in buildings \cite{als_Dieckmann1911}. Firstly it was tried to
improve safety by using non-flammable building materials. However,
the desaster at the Troquois Theater in Chicago with more than 500 
fatalities, where only the decoration burned, caused a rethinking. 
It was a starting point for studying the influences of
emergency exits and thus the dynamics of pedestrian streams
\cite{als_Dieckmann1911,als_Fischer1933}.

In recent years there were mainly two incidents including evacuations
which gained immense global attention. First there was the capsizing
of the Baltic Sea ferry MV Estonia (September 28, 1994, 852
casualties) \cite{Laur1997_tk} and then of course the terrorist
attacks of 9/11 (2,749 casualties).  Other prominent examples of the
possible tragic outcomes of the dynamics of pedestrian crowds are the
Hillsborough stadium disaster in Sheffield (April 15, 1989, 96
casualties) \cite{Taylor1990_tk}, 
the accident at Bergisel (December 4, 1999, 5 casualties)
\cite{WALD02}, the stampede in Baghdad (August 30, 2005, 1.011
casualties), the tragedy at the concert of ``The Who'' (December 3,
1979, 11 casualties) \cite{Johnson1987_tk} and -- very early -- the
events at the crowning ceremony of Tsar Nicholas II.  in
St.~Petersburg in May 1896 with 1,300 to 3,000 fatalities (sources vary
considerably) \cite{Schelajew2000_tk}. In the past tragic accidents
have happened frequently in Mecca during the Hajj (1990: 1,426, 1994:
270, 1997: 343, 1998: 107, 2001: 35, 2003: 14, 2004: 244, and 2006:
364 casualties).  
What stands out is that the initiating events are very diverse and
span from external human aggression (terrorism) over external physical
dangers (fire) and rumors to various shades of greedy behavior in
absence of any external danger.  


Many authors have pointed out that the results of
experts' investigations and the way the media typically reports about
an accident very often differ strongly
\cite{Quarantelli1960_tk,Sime1980_tk,Keating1982,Quarantelli2001_tk,Clarke2002,Mawson2005}.
The public discussion has a much greater tendency to
identify ``panic'' as cause of a disaster, while expert
commissions often conclude that there either was no panic
at all, or panic was merely a result of some other preceding
phenomenon.

The article first discusses the empirical basis of pedestrian dynamics
in Sec.~\ref{sec-emp}. Here we introduce the basic observables and
describe the main qualitative and quantitative results, focussing on
collective phenomena and the fundamental diagram.  It is emphasized
that even for the most basic quantities no consensus about the basic
behaviour has been reached.

In Sec.~\ref{sec_modelling} various model approaches that have been
applied to the description of pedestrian dynamics are reviewed.

Sec.~5 discusses more practical issues and gives a few examples for
applications to safety analysis. In this regard, prediction of
evacuation times is an important problem as often legal regulations
have to be fulfilled. Here commercial software tools are available.
A comparison shows that the results have to be interpreted with care.


\section{Empirical Results}
\label{sec-emp}

\subsection{Overview}

Pedestrians are three-dimensional objects and a complete description
of their highly developed and complicated motion sequence is
rather difficult. Therefore usually in pedestrian and 
evacuation dynamics the motion is treated as two-dimensional
by considering the vertical projection of the body.

In the following sections we review the present knowledge of empirical
results. These are relevant not only as basis for the
development of models, but also for applications like safety studies
and legal regulations.

We start with the phenomenological description of collective effects.
Some of these are known from everyday experience and will serve
as benchmark tests for any kind of modelling approach.
Any model that does not reproduce these effects is missing
some essential part of the dynamics.
Next the foundations of a quantitative description are laid by introducing
the fundamental observables of pedestrian dynamics. Difficulties
arise from different conventions and definitions. 
Then pedestrian dynamics in several simple scenarios
(corridor, stairs etc.) is discussed. Surprisingly even for these simple
cases no consensus about the basic quantitative properties exists.
Finally, more complex scenarios are discussed which are combinations
of the simpler elements. 
Investigations of scenarios like evacuations of large buildings or ships
suffer even more from lack of reliable quantitative and sometimes
even qualitative results.

\subsection{Collective Effects}
\label{sec-collective}

One of the reasons why the investigation of pedestrian dynamics is also
attractive for physicists is the large variety of interesting 
collective effects and self-organization phenomena that can be observed. 
These macroscopic effects reflect the individuals' microscopic interactions
and thus give also important information for any modelling approach.

\begin{itemize}

\item {\bf Jamming} 

Jamming and clogging typically occur for high densities at locations where
the inflow exceeds the capacity. Locations with reduced capacity are called 
{\em bottlenecks}. Typical examples are exits (Fig.~\ref{fig-ped-clogg})
or narrowings.
This kind of jamming phenomenon does not depend strongly on the microscopic
dynamics of the particles. Rather it is a consequence of an
exclusion principle: space occupied by one particle is not
available for others.

This clogging effect is typical for a bottleneck situation. It is
important for practical applications, especially evacuation
simulations.
\begin{figure}[h]
  \begin{center}
\includegraphics[width=0.45\textwidth]{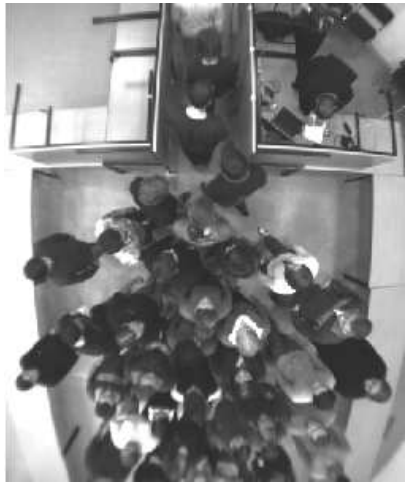}
    \caption{Clogging near a bottleneck. The shape of the clog is
discussed in more detail in Sec.~\ref{sec_theor-result}.
}
\label{fig-ped-clogg}
  \end{center}
\end{figure}

Other types of jamming occur in the case of counterflow where two
groups of pedestrians mutually block each other. This happens
typically at high densities and when it is not possible to turn
around and move back, e.g.\ when the flow of people is large.

\item {\bf Density waves}

  Density waves in pedestrian crowds can be generally characterised as
  quasi-periodic density variations in space and time.  A typical
  example is the movement in a densely crowded corridor (e.g. in
  subway-stations close to the density that causes a complete halt of
  the motion) where phenomena similar to stop-and-go vehicular traffic
  can be observed, e.g. density fluctuations in longitudinal direction
  that move backwards (opposite to the movement direction of the
  crowd) through the corridor.  More specifically, for the situation
  on the Jamarat Bridge in Makkah (during the Hajj pilgrimage 2006)
  stop-and-go waves have been reported. At densities of 7 persons per
  m$^2$ upstream moving stop-and-go waves of period 45~s have been
  observed that lasted for 20 minutes \cite{Helbing2007}.
  Fruin reports, that ``at occupancies of about 7 persons
  per square meter the crowd becomes almost a fluid mass. Shock
  waves can be propagated through the mass sufficient to lift people
  of their feet and propel them distances of 3~m (10~ft) or
  more.'' \cite{Fruin1993}. 

\item {\bf Lane formation}

In counterflow, i.e.\ two groups of people moving in opposite directions, 
(dynamically varying) lanes are formed where people move in just one
direction \cite{Oeding1963,Navin1969,Yamori1998}. In this way, strong 
interactions with
oncoming pedestrians are reduced which is more comfortable and allows 
higher walking speeds.

The occurrence of lane formation does not require a preference
of moving on one side. It also occurs in situations without
left- or right-preference. However, cultural differences
for the preferred side have been observed.
Although this preference is not essential for the phenomenon itself,
it has an influence on the kind of lanes formed and their order.

Several quantities for the quantitative characterization
of lane formation have been proposed.
Yamori \cite{Yamori1998} has introduced a band index
which is basically the ratio of pedestrians in lanes to their total
number. 
In \cite{Burstedde2001a} a characterization of lane formation
through the (transversal) velocity profiles at fixed positions
has been proposed.
Lane formation has also been predicted to occur in colloidal mixtures 
driven by an external field \cite{DzubiellaHL02A,ChakrabartiDL04A,RexL07A}.
Here an order parameter
$\phi = \frac{1}{N}\left\langle \sum_{j=1}^N \phi_j\right\rangle$
has been introduced
where $\phi_j=1$ if the lateral distance to all other particles of
the other type is larger than a typical density-dependent length scale
and $\phi_j=0$ otherwise.

The number of lanes can vary considerably with the total width of the 
flow. Fig.~\ref{pede-WJT_Koeln} shows a street in the city center of
Cologne during the World Youth Day in Cologne (August 2005) where
two comparatively large lanes have been formed.

The number of lanes usually is not constant and might change in time,
even if there are relatively small changes in density.
The number of lanes in opposite directions is not always identical.
This can be interpreted as a sort of spontaneous symmetry breaking. 
\begin{figure}[thb]
        \begin{center}
  \includegraphics[width=.75\textwidth]{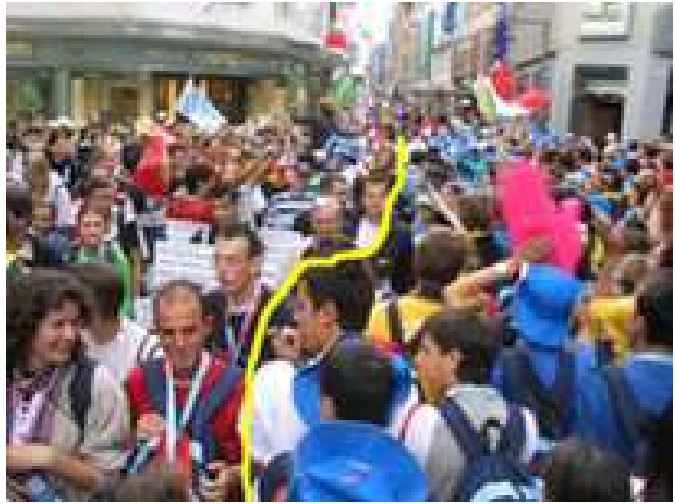}
  \caption{The ``Hohe Stra{\ss}e'' in Cologne during World Youth Day 2005. 
The yellow line is the border of the two walking directions.
}
  \label{pede-WJT_Koeln}
  \end{center}
\end{figure}

Quantitative empirical studies of lane formation are rare. 
Experimental result have been reported in \cite{KretzGKMS06A} where two 
groups with varying relative sizes had to pass each other in a corridor 
with a width of 2~m.  
On one hand, similar to \cite{Yamori1998}  a variety of
different lane patterns were observed, ranging from 2 to 4 lanes.
On the other hand,
in spite of this complexity surprisingly large flows could be
measured: the sum of (specific) flow and counterflow was between 1.8
and 2.8 persons per meter and second and exceeded the specific flow
for one-directional motion ($\approx$1.4~P/ms).

\item {\bf Oscillations} 

In counterflow at bottlenecks, e.g.\ doors, one can sometimes observe 
oscillatory changes of the direction of motion. Once a
pedestrian is able to pass the bottleneck it becomes easier for
others to follow in the same direction until somebody is able to
pass (e.g.\ through a fluctuation) the bottleneck in the opposite
direction.

\item {\bf Patterns at intersections}

At intersections various collective
patterns of motion can be formed. A typical example are short-lived
roundabouts which make the motion more efficient. Even if these are
connected with small detours the formation of these patterns can be
favourable since they allow for a ``smoother'' motion.

\item {\bf Emergency situations, ``panic''} 

In emergency situations various collective phenomena have been
reported that have sometimes misleadingly been attributed to 
{\em panic behaviour}.
However, there is strong evidence that this is not the case.
Although a precise accepted definition of {\em panic} is missing,
usually certain aspects are associated with this concept \cite{Keating1982}. 
Typically ``panic'' is assumed to occur in situations where people 
compete for scarce or dwindling resources (e.g.\ safe space
or access to an exit) which leads to selfish, asocial or even completely
irrational behaviour and contagion that affects large groups.
A closer investigation of many crowd disasters 
has revealed that most of the above characteristics have played
almost no role and most of the time have not
been observed at all (see e.g.\ \cite{Johnson1987_tk}). 
Often the reason for these accidents is much simpler, e.g.\ in several 
cases the capacity of the facilities was too small for the actual 
pedestrian traffic, e.g. Luschniki Stadium Moskau (October 20, 1982), 
Bergisel (December 4, 1999), pedestrian bridge Kobe (Akashi) (July 21, 2001) 
\cite{Tsuji03}. Therefore the
term ``panic'' should be avoided, {\em crowd disaster} being a more
appropriate characterisation. 
Also it should be kept in mind that in dangerous situations it is {\em not} 
irrational to fight for resources (or your own life), if everybody else 
does this \cite{Mintz1951,Coleman1990}. Only from the outside this behavior
is perceived as irrational since it might lead to a catastrophe
\cite{Sime1980_tk}. The latter aspect is therefore better described as
{\em non-adaptive behvaiour}.

We will discuss these issues in more detail in Sec.~\ref{sub-evac-emp}.

\end{itemize}

\subsection{Observables}
\label{sec-observables}

Before we review experimental studies in this section, the 
commonly used observables are introduced.

The flow $J$ of a pedestrian stream gives the number of pedestrians
crossing a fixed location of a facility per unit of time. 
Usually it is taken as a scalar quantity since only the flow
normal to some cross-section is considered.
There are various methods to measure the flow. The most natural
approach is to determine the times $t_i$ at which pedestrians passed a
fixed measurement location. The time gaps $\Delta t_i=t_{i+1}-t_{i}$
between two consecutitive pedestrians $i$ and $i+1$ are directly
related to the flow
\begin{equation}
J = \frac{1}{\langle\Delta t\rangle} 
\qquad \mbox{with} \qquad 
\langle\Delta t\rangle=\frac{1}{N}\sum_{i=1}^{N}(t_{i+1}-t_{i})
=\frac{t_{N+1}-t_{1}}{N}\,.
\label{FLOW_ALS}
\end{equation}

Another possibility to measure the flow of a pedestrian stream is borrowed 
from fluid dynamics.
The flow through a facility of width  $b$ determined by the
average density $\rho$ and the
average speed $v$ of a pedestrian stream as
\begin{equation}
J = {\rho\;v}\;b = J_s b\,.
\label{FLUIDFLOW_ALS}
\end{equation}
where the {\em specific flow}\footnote{In strictly one-dimensional
motion often a line density (dimension: 1/length) is used. Then the
{\em flow} is given by $J=\rho v$.}
\begin{equation}
J_s = {\rho\;v} 
\label{hydroRel}
\end{equation}
gives the flow per unit-width. This relation is also known
as {\em hydrodynamic relation}. 

There are several problems concerning the way how velocities,
densities or time gaps are measured and the conformance of the two
definitions of the flow. The flow according to eq.~(\ref{FLOW_ALS}) is
usually measured as a mean value over time at a certain location while
the measurement of the density in eq.~(\ref{FLUIDFLOW_ALS}) is connected
with an instantaneous mean value over space. This can lead to a bias
caused by the underestimation of fast moving pedestrians at the
average over space compared to the mean value of the flow over time at
a single measurement line, see the discussion for vehicular traffic
e.g.\ in \cite{Leutzbach1988,HelbingRMP,Kerner2004}.  Furthermore most
experimental studies measuring the flow according to equation
(\ref{FLUIDFLOW_ALS}) combine for technical reasons a \emph{average}
velocity of a single pedestrian over time with an \emph{instantaneous}
density.  To ensure a correspondence of the mean values the average
velocity of all pedestrians contributing to the density at a certain
instant has to be considered.  However this procedure is very time
consuming and not realised in practice up to now.  Moreover the fact
that the dimension of the test section has usually the same order of
magnitude as the extent of the pedestrians can influence the averages
over space. These all are possible factors why different measurements
can differ in a large way, see discussion in Sec.~\ref{FUNDDIA}.

Another way to quantify the pedestrian load of facilities has been
proposed by Fruin \cite{Fruin1971}. The ``pedestrian area module'' is given 
by the reciprocal of the density.
Thompson and Marchant \cite{Thompson1994_cr} introduced the so-called 
``inter-person distance'' $d$, which is measured between centre coordinates 
of the assessing and obstructing persons. According to the ``pedestrian area
module'' Thompson and Marchant call $\sqrt{\frac{1}{\rho}}$ the ``average
inter-person distance'' for a pedestrian stream of evenly spaced 
persons \cite{Thompson1994_cr}.
An alternative definition is introduced in \cite{als_helbing07a} where
the local density is obtained by averaging over a circular region
of radius $R$, 
\begin{equation}
\rho(\vec{r},t) = \sum_j f(\vec{r}_j(t)-\vec{r}),
\end{equation}
where $\vec{r}_j(t)$ are the positions of the pedestrians $j$ in the
surrounding of $\vec{r}$ and $f(...)$ is a Gaussian, distance-dependent
weight function.  

In contrast to the density definitions above, Predtechenskii and
Milinskii \cite{Predtetschenski1971} consider the ratio of
the sum of the projection area $f_j$ of the bodies and
the total area of the pedestrian stream $A$, defining the (dimensionless)
density $\tilde{\rho}$ as
\begin{equation}
    \tilde{\rho} = \frac{\sum_j{f_j}}{A}\,,
\end{equation}
a quantity known as \emph{occupancy} in the context of
vehicular traffic.
Since the projection area $f_j$ depends strongly on the type of
person (e.g.\ it is much smaller for a child than an adult), the densities
for different pedestrian streams
consisting of the same number of persons and the same stream area 
can be quite different.

Beside technical problems due to camera distortions and camera
perspective there are several conceptual problems, like the
association of averaged with instantaneous quantities, the necessity
to choose an observation area in the same order of magnitude as the
extent of a pedestrian together with the definition of the density of
objects with non zero extent and much more. A detailed analysis how
the way of measurement influences the relations is necessary but still
lacking.

\subsection{Fundamental Diagram}
\label{FUNDDIA}

The fundamental diagram describes the empirical relation between 
density $\rho$ and flow $J$. The name already indicates its importance 
and naturally it has been the subject of many investigations. 
Due to the hydrodynamic relation (\ref{hydroRel}) there are three
equivalent forms: $J_s(\rho)$, $v(\rho)$ and $v(J_s)$.
In applications the relation is a basic input for engineering methods
developed for the design and dimensioning of pedestrian facilities
\cite{als_PRED78ENG,Fruin1971,Nelson2002}. Furthermore it is a 
quantitative benchmark for models of pedestrian dynamics
\cite{als_SCHR02b,Kirchner2004,als_Hoogendoorn02b,Seyfried2006c}. 

In this section 
we will concentrate on planar facilities like sidewalks, corridors or halls.
For various facilities like floors, stairs or ramps
the shape of the diagrams differ, but in general it is assumed that the
fundamental diagrams for the same type of facilities but different widths
merge into one diagram for the specific flow $J_s$. In first order
this is confirmed by measurements on different widths
\cite{als_HANK58,Oeding1963,als_OLD68,Navin1969}.  
However, Navin and Wheeler observed in narrow sidewalks more orderly 
movement leading to slightly higher specific flows than for
wider sidewalks \cite{Navin1969}. A natural lower bound for the
independence of the specific flow from the width is given by the body
size and the asymmetry in movement possibilities of the human body.
Surprisingly Kretz et al.\ found an increase of the specific flow for
bottlenecks with $b \le 0.7$~m \cite{als_kretz-06c}.  This will be
discussed in more detail later. For the following discussion we assume
facility widths larger than $b=0.6$~m and use the most common
representations $J_s(\rho)$ and $v(\rho)$.

\begin{figure}[thb]
  \begin{center}
    \includegraphics[width=0.46\textwidth]{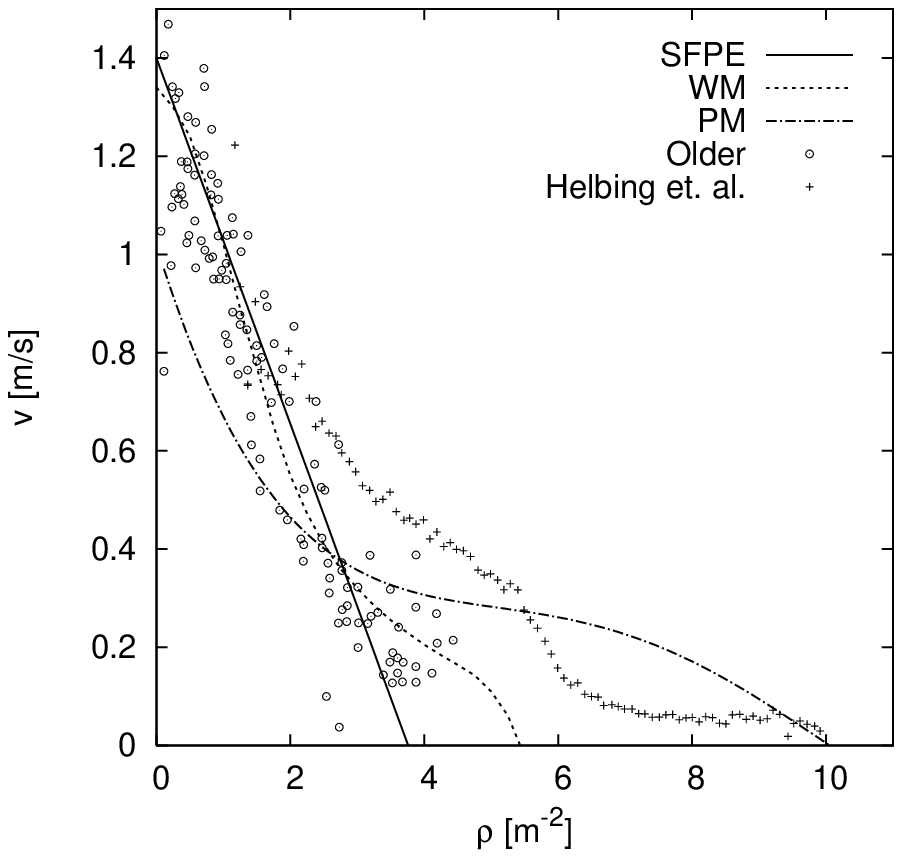}
    \qquad
    \includegraphics[width=0.46\textwidth]{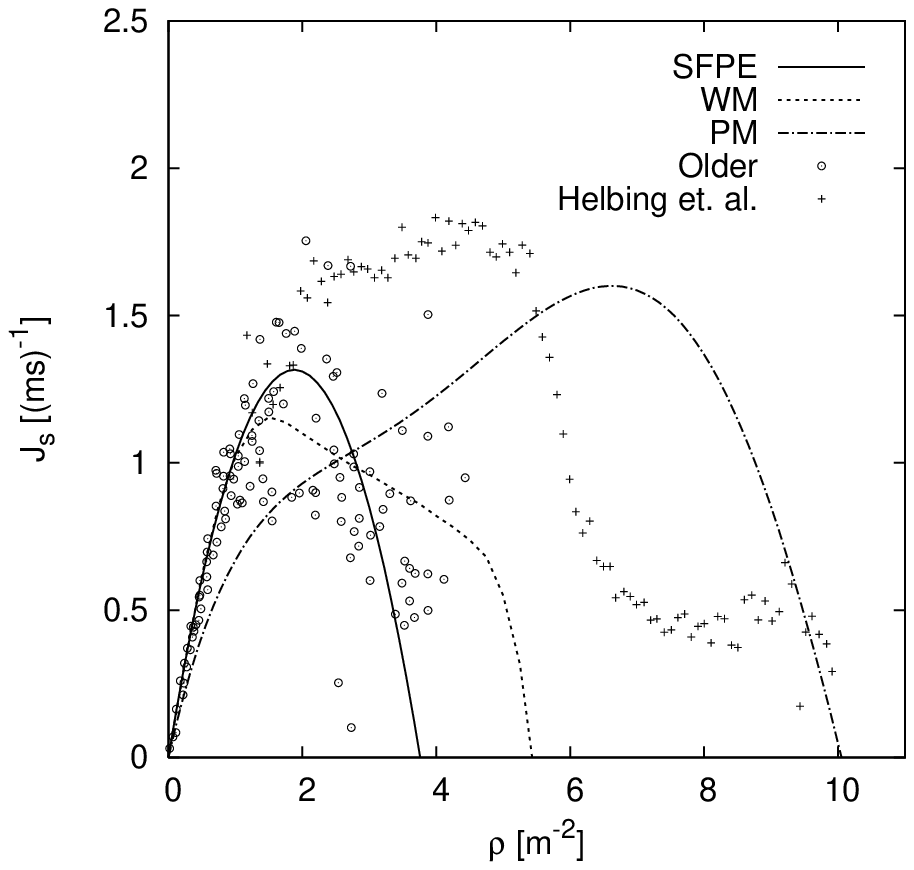}
    \caption{Fundamental diagrams for pedestrian movement in planar facilities.
      The lines refer to specifications according to planing
      guidelines (SFPE Handbook \cite{Nelson2002}), Predtechenskii and
      Milinskii (PM) \cite{als_PRED78ENG}, Weidmann (WM)
      \cite{Weidmann1993}). Data points give the range of experimental
      measurements (Older \cite{als_OLD68} and Helbing
      \cite{als_helbing07a}).
      \label{RHO-JS}
}
  \end{center}
\end{figure}

Fig.~\ref{RHO-JS} shows various fundamental diagrams used in planing
guidelines and measurements of two selected empirical studies
representing the overall range of the data. The comparison reveals
that specifications and measurements disagree considerably. In
particular the maximum of the function giving the capacity
$J_{s,\text{max}}$ ranges from $1.2\;$(ms)$^{-1}$ to
$1.8\;$(ms)$^{-1}$, the density value where the maximum flow is
reached $\rho_c$ ranges from $1.75\;$m$^{-2}$ to $7\;$m$^{-2}$ and,
most notably, the density $\rho_0$ where the velocity approaches
zero due to overcrowding ranges from $3.8\;$m$^{-2}$ to
$10\;$m$^{-2}$.

Several explanations for these deviations have been suggested,
including cultural and population differences
\cite{als_MORR91,als_helbing07a}, differences between uni- and
multidirectional flow \cite{Navin1969,als_pushkarev75,als_LAM03},
short-ranged fluctuations \cite{als_pushkarev75}, influence of
psychological factors given by the incentive of the movement
\cite{als_PRED78ENG} and, partially related to the latter, the type of
traffic (commuters, shoppers) \cite{Oeding1963}.

It seems that the most elaborate fundamental diagram is given by
Weidmann who collected 25 data sets. An examination of the data which
were included in Weidmann's analysis shows that most measurements with
densities larger then $\rho=1.8\;$m$^{-2}$ are performed on
multidirectional streams
\cite{als_OLD68,Navin1969,Oeding1963,als_OFL72,als_POL83}.  But also
data gained by measurements on strictly unidirectional streams has
been considered \cite{als_HANK58,Fruin1971,als_VIRK94b}. Thus
Weidmann neglected differences between uni- and multidirectional flow
in accordance with Fruin, who states in his often cited book
\cite{Fruin1971} that the fundamental diagrams of multidirectional 
and unidirectional flow differ only slightly.  
This disagrees with results of Navin and Wheeler \cite{Navin1969} and 
Lam et al.\ \cite{als_LAM03} who
found a reduction of the flow in dependence of directional imbalances.
Here lane formation in bidirectional flow has to be considered. 
Bidirectional pedestrian flow includes
unordered streams as well as lane-separated and thus quasi-unidirectional
streams in opposite directions. A more detailed discussion and data can
be found in \cite{Navin1969,als_pushkarev75,als_LAM03}. A surprising 
finding is that the sum of flow and counterflow in corridors is larger 
than the unidirectional flow and for equally distributed loads it can be 
twice the unidirectional flow  \cite{KretzGKMS06A}.

Another explanation is given by Helbing et al.\ \cite{als_helbing07a}
who argue that cultural and population differences are responsible
for the deviations between Weidmann and their data. In contrast to
this interpretation the data of Hanking and Wright \cite{als_HANK58}
gained by measurements in the London subway (UK) are in good agreement
with the data of Mori and Tsukaguchi \cite{als_MORI_87} measured in
the central business district of Osaka (Japan), both on strictly
uni-directional streams. This brief discussion clearly shows that up to
now there is no consensus about the origin of the discrepancies
between different fundamental diagrams and how one can explain the
shape of the function.

However, all diagrams agree in one characteristic: velocity decreases
with increasing density. As the discussion above indicates
there are many possible reasons and causes for the velocity reduction.
For the movement of pedestrians along a line a linear relation between
speed and the inverse of the density was measured in \cite{als_SEY05a}.
The speed for walking pedestrians depends also linearly on the step
size \cite{Weidmann1993} and the inverse of the density can be regarded as
the required length of one pedestrian to move. Thus it seems that smaller
step sizes caused by a reduction of the available space with
increasing density is, at least for a certain density region, one cause
for the decrease of speed. However, this is only a starting point
for a more elaborated modeling of the fundamental diagram.

\subsection{Bottleneck Flow} 
\label{BCK}

The flow of pedestrians through bottlenecks shows a rich variety
of phenomena, e.g.\ the formation of lanes at the entrance to the 
bottleneck
\cite{als_Hoogendoorn03a,als_Hoogendoorn05a,als_kretz-06c,als_SEY07a},
clogging and blockages at narrow bottlenecks
\cite{als_Dieckmann1911,als_PRED78ENG,als_Mueller81,Muir1996,als_HELB05a,als_kretz-06c}
or some special features of bidirectional bottleneck flow
\cite{als_HELB05a}.  Moreover, the estimation of bottleneck capacities by
the maxima of fundamental diagrams is an important tool for the
design and dimensioning of pedestrian facilities.

\subsubsection{Capacity and bottleneck width}

One of the most important practical questions is how the capacity of
the bottleneck increases with rising width. Studies of this dependence
can be traced back to the beginning of the last century
\cite{als_Dieckmann1911,als_Fischer1933} and are up to now discussed
controversially. As already mentioned in the context of the fundamental
diagram there are multiple possible influences on pedestrian flow and
thus on the capacity.  In the following the major findings are
outlined, demonstrating the complexity of the system and documenting
a controversial discussion over one hundred years.

\begin{figure}[thb]
  \begin{center}
    \includegraphics[width=0.75\textwidth]{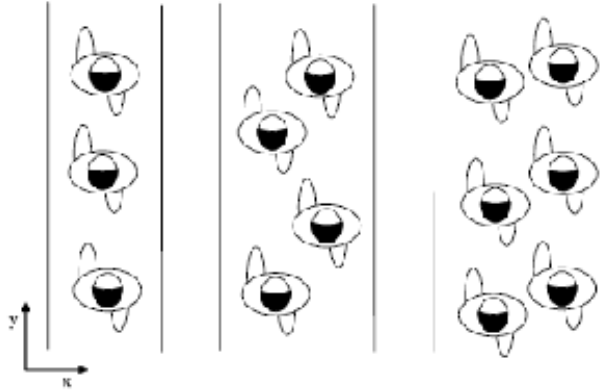}
    \caption{A sketch of the zipper effect with continuously increasing 
      lane distances in $x$: The distance in the walking
      direction decreases with increasing lateral distance. 
      Density and velocities are the same in all cases, but the flow
      increases continuously with the width of the section.
\label{ZIPPER}}
  \end{center}
\end{figure}

At first sight, a stepwise increase of capacity with the width appears
to be natural if lanes are formed. 
For independent lanes, where pedestrians in one lane are not
influenced by those in others, 
the capacity increases only if an additional lane can be formed. 
This is reflected in the stepwise enlargement of exit width which is up 
to now a requirement of several building codes and design
recommendations, see e.g.\ the discussion in \cite{als_Pauls06a} for the
USA and GB and \cite{als_MVSTAETT} for Germany. E.g. the German building 
code requires an exit width (e.g.\ for a door) to be 90~cm at least and 
60~cm for every 200 persons.
Independently from
this simple lane model Hoogendoorn and Daamen
\cite{als_Hoogendoorn03a,als_Hoogendoorn05a} measured by a laboratory
experiment the trajectories of pedestrians passing a bottleneck. The
trajectories show that inside a bottleneck the formation of lanes
occurs, resulting from the zipper effect during entering the
bottleneck. Due to the zipper effect, a self-organization phenomenon
leading to an optimization of the available space and velocity, the
lanes are not independent and thus do not allow passing (Fig.~\ref{ZIPPER}). 
The empirical results of \cite{als_Hoogendoorn03a,als_Hoogendoorn05a} 
indicate a distance between lanes of $d \approx 0.45\;$m, independent
of the bottleneck width $b$, implying a stepwise increase of capacity. 
However, the investigation was restricted to two values ($b=1.0\;$m 
and $b=2.0\;$m) of the width.

In contrast, the study \cite{als_SEY07a} considered more values of the
width and found that the lane distance increases continuously as
illustrated in Fig.~\ref{ZIPPER}.
Moreover it was shown that the continuous increase of the lane 
distance leads to a very weak dependence of the density and velocity 
inside the bottleneck on its width. Thus in reference to 
eq.~(\ref{FLUIDFLOW_ALS}) the flow does not necessarily depend on the 
number of lanes. This is consistent with common guidelines and 
handbooks\footnote{One exception is the German MVSt{\"a}ttV 
\cite{als_MVSTAETT}, see above.} which assume that the capacity is a linear
function of the width \cite{Fruin1971,als_PRED78ENG,Weidmann1993,Nelson2002}.  
It is given by the maximum of the fundamental diagram and in reference
to the specific flow concept introduced in Section~\ref{sec-observables}, 
eqs.~(\ref{FLUIDFLOW_ALS}), (\ref{hydroRel}), the maximum grows linearly
with the facility width. To find a conclusive judgement on the
question if the capacity grows continuously with the width the results
of different laboratory experiments
\cite{als_Mueller81,Muir1996,als_NAGAI06,als_kretz-06c,als_SEY07a}
are compared in \cite{als_SEY07a}.

\begin{figure}[thb]
  \begin{center}
    \includegraphics[width=0.85\textwidth]{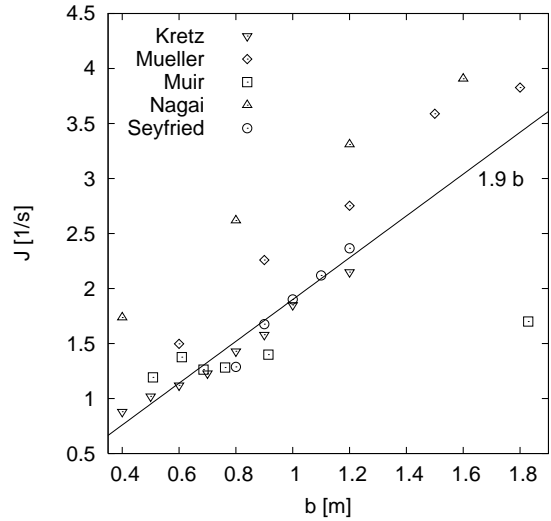}
    \caption{Influence of the width of a bottleneck on the flow. 
      Experimental data \cite{als_Mueller81,Muir1996,als_NAGAI06,als_SEY07a}
      of different types of bottlenecks and initial
      conditions. All data are taken under laboratory conditions where
      the test persons are advised to move normally. 
\label{J-B-EMP}}
  \end{center}
\end{figure}

\begin{figure}
  \centering \subfigure[Kretz]{\label{ESET:suba}
    \includegraphics[width=.25\textwidth]{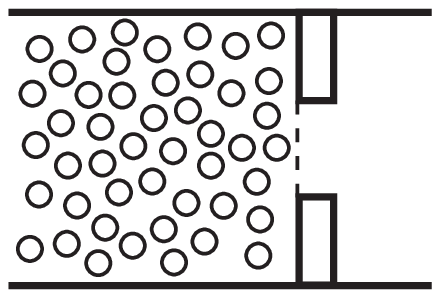}}
  \hspace{20.pt} \subfigure[Muir]{\label{ESET:subb}
    \includegraphics[width=.25\textwidth]{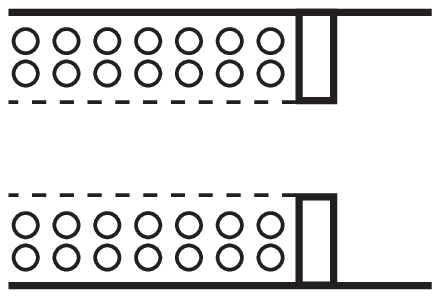}}
  \hspace{20.pt} \subfigure[M\"uller]{\label{ESET:subc}
    \includegraphics[width=.25\textwidth]{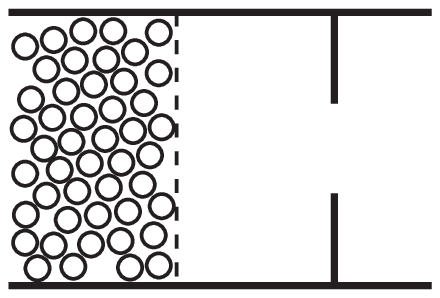}}
  \hspace{20.pt} \subfigure[Nagai]{\label{ESET:subd}
    \includegraphics[width=.25\textwidth]{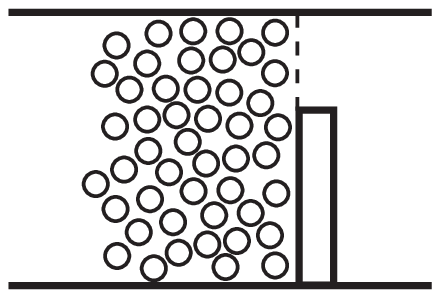}}
  \hspace{20.pt} \subfigure[Seyfried]{\label{ESET:sube}
    \includegraphics[width=.25\textwidth]{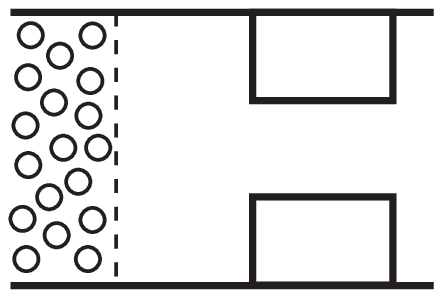}}
  \caption{Outlines of the experimental arrangements under which the 
data shown in Figure \ref{J-B-EMP} were taken.}
  \label{ESET:sub}
\end{figure} 

In the following we discuss the data of flow measurement collected in
Fig.~\ref{J-B-EMP}. The corresponding setups are sketched in
Fig.~\ref{ESET:sub}.  First one has to note that all presented data are
taken under laboratory conditions where the test persons are advised
to move normally. The data by Muir et al.\ \cite{Muir1996}, who
studied the evacuation of airplanes (see Fig.~\ref{ESET:subb}), seem 
to support the stepwise increase of the flow with the width. They show 
constant flow values for $b > 0.6$~m. But the independence of the
flow over the large range from $b=0.6$~m to $b=1.8$~m indicates that 
in this special setup the flow is not restricted by the bottleneck
width. 
Moreover it was shown in \cite{als_SEY07a} by determination
of the trajectories that the distance between lanes changes continuously,
invalidating the basic assumption leading to a stepwise increasing flow.  
Thus all collected data for flow measurements
in Fig.~\ref{J-B-EMP} are compatible with a continuous and
almost linear increase with the bottleneck width for $b>0.6\;m$.

Surprisingly the data in Fig.~\ref{J-B-EMP} differ considerably in the
values of the bottleneck capacity.  In particular the flow
values of Nagai \cite{als_NAGAI06} and M\"uller \cite{als_Mueller81}
are much higher than the maxima of empirical fundamental
diagrams, see Sec.~\ref{FUNDDIA}. The influence of ``panic'' or
pushing can be excluded since in all experiment the participants
were instructed to move normally.  The
comparison of the different experimental setups (Fig.~\ref{ESET:sub})
shows that the exact geometry of the bottleneck is of only minor
influence on the flow while a high initial density in front of the
bottleneck can increase the resulting flow values. This is confirmed
by the study of Nagai et al., see Figure 6 in \cite{als_NAGAI06}.
There it is shown that for $b=1.2\;$m the flow grows from
$J=1.04\;$s$^{-1}$ to $3.31\;$s$^{-1}$ when the initial density is
increased from $0.4\;$m$^{-2}$ to $5\;$m$^{-2}$.

The linear dependence of the flow on the width has a natural
limitation due to the non-zero body-size and the asymmetry given by
the sequence of movement in steps. The moving of pedestrians through
bottlenecks smaller than the shoulder width, requires a rotation of
the body. Surprisingly Kretz et al.  found in their
experiment \cite{als_kretz-06c} that the specific flow $J_s$ increases
if the width decreases from $b=0.7\;$m to $b=0.4\;$m. 

\subsubsection{Connection between bottleneck flow and fundamental diagrams}

An interesting question is how the bottleneck flow is connected to
the fundamental diagram. General results for driven diffusive systems
\cite{Popkov1999A} show that boundary conditions only \emph{select}
between the states of the undisturbed system instead of creating
completely different ones. Therefore it is surprising
that the measured maximal flow at bottlenecks can exceed the maximum of 
the empirical fundamental diagram.
These questions are related to the common jamming criterion. Generally 
it is assumed that a jam occurs if the incoming flow exceeds the capacity 
of the bottleneck. In this case 
one expects the flow through the bottleneck to continue with the capacity 
(or lower values).

The data presented in \cite{als_SEY07a} show a more complicated
picture. While the density in front of the bottleneck amounts to $\rho
\approx 5.0~(\pm 1) \;$m$^{-2}$, the density inside the bottleneck
tunes around $\rho \approx 1.8\,$m$^{-2}$. The
observation that the density inside the bottleneck is lower than in
front of the bottleneck is consistent with measurements of Daamen and
Hoogendoorn \cite{als_Hoogendoorn06c} and the description given by
Predtechenskii and Milinskii in \cite{als_PRED78ENG}. The latter
assumes that in the case of a jam the flow through the bottleneck is
determined by the flow in front of the bottleneck. The density inside
the jam will be higher than the density associated with the capacity.
Thus the reduced flow in front of the bottleneck causes a flow through
the bottleneck smaller than the bottleneck capacity. Correspondingly
the associated density is also smaller than that at capacity. But the
discussion above can not explain why the capacities measured at
bottlenecks are significant higher than the maxima of empirical
fundamental diagrams and cast doubts on the common jamming criterion.
Possible unconsidered influences are stochastic flow fluctuations, 
non-stationarity of the flow, flow
interferences due to the necessity of local organization or
changes of the incentive during the access into the bottleneck.  

\subsubsection{Blockages in competitive situations}

As stated above all data collected in figure \ref{J-B-EMP} are gained
by runs where the test persons were instructed to move normally. By
definition a bottleneck is a limited resource and it is possible that
under competitive situation pedestrian flow through bottlenecks is
different from the flow in normal situations.  
One qualitative difference to normal situations is the
occurance of blockages. Regarding the term `panic' one has
to bear in mind that for the occurance of blockages
some kind of reward is essential while the emotional
state of the test persons is not. This was a result of
a very interesting and often cited study by Mintz \cite{Mintz1951}.
First experiments with real pedestrians have been performed by Dieckmann 
\cite{als_Dieckmann1911} in 1911 as reaction to many fatalities in theater 
fires at the end of the 19th century.
In these small scale experiments test persons were instructed to go through
great trouble to pass the door as fast as possible. Even in the first
run he observed a stable ``wedging''. 
In \cite{als_PRED78ENG} it is 
described how these obstruction occurs due to the formation of arches
in front of the door under high pressure. 
This is very similar to the well-known phenomenon of {\em arching} 
occuring in the flow of granular materials through narrow 
openings \cite{Wolf1996}.

Systematic studies including the influence of the shape and width of the
bottleneck and the comparison with flow values under normal situations
have been performed by M\"uller and Muir et al.\ 
\cite{als_Mueller81,Muir1996}. M\"uller found that funnel-like
geometries support the formation of arches and thus blockages. For the
further discussion one has to distinguish between temporary blockages
and stable blockages leading to a zero flow. For the setup sketched in
Fig.~\ref{ESET:subc} M\"uller found that temporary blockages occur
only for $b < 1.8\,$m. For $b \leq 1.2\,$m the flow shows strong
pulsing due to unstable blockages.  Temporal disruptions of the flow
establish for $b\leq 1.0\;$m.  In comparison to normal situations the
flow is higher and in general the occurrence of blockages decrease
with width. However a surprising result is that for narrow bottlenecks
increasing the width can be counterproductive since it also increases 
the probability of blockages. Muir et al. for example note that
in their setup (Fig.~\ref{ESET:subb}) the enlargement of the
width from $b=0.5\;$m to $b=0.6\;$m leads to an increase of temporary
blockages. The authors explain this by differences in the perception
of the situation by the test persons. While the smaller width is
clearly passable only for one person the wider width may lead to the
perception that the bottleneck is sufficiently wide to allow two persons
to pass through. How many people have direct access to the bottleneck is
clearly influenced by the width of the corridor in front of the
bottleneck. Also M\"uller found hints that flow under competitive
situations did not increase in general with the bottleneck width. He
notes an optimal ratio of $0.75:1$ between the bottleneck width and the
width of the corridor in front of the bottleneck.

To reduce the occurrence of blockages and thus evacuation times,
Helbing et al.\ \cite{Helbing2000,Helbing2002,Kirchner2002b} suggested to 
put a column (asymmetrically) in front of a bottleneck.
It should be emphasized that this theoretical prediction was made
under the assumption that the system parameters, i.e.\ the basic behaviour
of the pedestrians, does not change in the presence of the column. 
This is highly questionable in real situations where the columns can
be perceived as an additional obstacle or even make it difficult to
find the exit.
In experiments \cite{als_HELB05a} an increase of the flow of about 
$30\%$ has been observed for a door with $b=0.82\;$m.  
But this experiment was only performed for one width and the discussion 
above indicates the strong influence of the specific setup used.
Independent of this uncertainty this concept is limited, as the
occurrence of stable arches, to narrow bottlenecks.  In practice narrow
bottlenecks are not suitable for a large number of people and an opening
in a room has also other important functionalities, which would be
restricted by a column.

Another surprising finding is the observation that the total flow at
bottlenecks with bidirectional movement is higher than it is for
unidirectional flows \cite{als_HELB05a}. 

\subsection{Stairs} 

In most evacuation scenarios stairs are important elements that are a
major determinant for the evacuation time. Due to their physical
dimension which is often smaller than other parts of a building or due
to a reduced walking speed, stairs generally have to be considered as
bottlenecks for the flow of evacuees. For the movement on stairs, just
as for the movement on flat terrain, the fundamental diagram is of
central interest.  Compared to the latter one there are more degrees
of freedom, which influence the fundamental diagram:
\begin{list}{-}{}
\setlength{\itemsep}{1pt}
\setlength{\parskip}{0pt}
\item One has to distinguish between upward and downward movement.
\item The influence of riser height and tread length 
(which determine the incline) has to be taken into account.
\item For upward motion exhaustion effects lead to a strong time
  dependence of the free speed.
\end{list}
It is probably a consequence of the existence of a continuum of
fundamental diagrams in dependence of the incline that there are no
generally accepted fundamental diagrams for the movement on stairs.
However, there are studies on various details --- mostly the free speed
--- of motion on stairs in dependence of the incline
\cite{Fujiyama2004_tk,Fujiyama2004b_tk,Graat1999,Fruin1971},
conditions (comfortable, normal, dangerous)
\cite{Predtetschenski1971}, age and sex
\cite{Fruin1971}, tread width \cite{Frantzich1996}, and the
length of a stair \cite{Kretz2007}; and in consideration of various
disablements \cite{Boyce1999_tk}.

In addition there are some compilations or ``meta studies'': Graat
\cite{Graat1999} compiled a list of capacity measurements and Weidmann
\cite{Weidmann1993} built an average of 58 single studies and found
an average for the horizontal upstairs speed --- the speed when the 
motion is projected to the horizontal level --- of 0.610~m/s.

Depending on the various parameters of aforesaid studies, those
studies report horizontal upward walking speeds varying over a wide range
from 0.391 to 1.16~m/s. Interestingly on one and the
same short stairs it could be observed \cite{Kretz2007} that people
on average walked faster up- than downwards.

There is also a model where the upstairs speed is calculated from the
stair geometry (riser and tread) \cite{Templer1992_tk} and an
empirical investigation of the collision avoidance behavior on stairs
\cite{Fujiyama2006_tk}.

On stairs (up- as well as downward) people like to put their hand on
the handrail, i.e. they tend to walk close to walls, even if there is
no counterflow. This is in contrast to movement on flat terrain, where
at least in situations of low density there is a tendency to keep some
distance from walls.

The movement on stairs is typically associated with a reduction of the
walking speed. For upward motion this follows from the increased
physical effort upward motion requires. This has two aspects, first
there is the physical potential energy that a pedestrian has to
supply if he wants to rise in height, second the motion process itself
is more exertive - the leg has to be lifted higher - than during
motion on a level, even if this motion process is executed only on the spot. 
Concerning the potential energy there is no comparable effect
for people going downstairs. But still one can observe jams forming at
the upper end of downstairs streams. These are due to the slight
hesitation that occurs when pedestrians synchronize their steps with
the geometry of the (down-)stairs ahead. Therefore the bottleneck
character of downstairs is less a consequence of the speed on the
stairs itself and more of the transition from planar to downward
movement, at least as long as the steps are not overly steep.

\subsection{Evacuations: Empirical results}
\label{sub-evac-emp}

Up to now this section has focussed on empirical results for
pedestrian motion in rather simple scenarios. As we have seen
there are many open questions where no consensus has been reached,
sometimes even about the qualitative aspects. This becomes even more
relevant for full-scale descriptions of evacuations from large buildings
or cruise ships. These are typically a combination of many of the simpler
elements. Therefore a lack of reliable information is not surprising. 
In the following we will discuss several complex scenarios in
more detail.


\subsubsection{Evacuation Experiments}

In the case of an emergency, the movement of a crowd usually is more
straightforward than in the general case. Commuters in a railway
station, for example, or visitors of a building might have complex
itineraries which are usually represented by origin-destination
matrices. In the case of an evacuation, however, the aims and routes
are known and usually the same, i.e. the exits and the egress routes.
This is the reason why an evacuation process is rather strictly
limited in space and time, i.e.\ its beginning and end are well-defined
(sound of the alarm, initial position of all persons, safe areas, final
position of all persons, and the time, the last person reaches the safe
area). When all people have
left a building or vessel and reached a safe area (or the lifeboats or
liferafts), then the evacuation is finished. Therefore, it is also
possible to perform evacuation trials and measure overall evacuation
times. Before we go into details, we will clarify three different
aspects of data on evacuation processes: (1) the definition and parts
of evacuation time, (2) the different sources of data, and (3) the
application of these data.

Concerning the evacuation time five different phases can be 
distinguished \cite{Purser2001,Hamacher2002,MSC-Circ.10332002}: 
(1) detection time, (2) awareness time, (3) decision time, 
(4) reaction time, and (5) movement time.  
In IMO's regulations \cite{MSC-Circ.10332002,MSC-Circ.11662005}, the first 
four are grouped together into \emph{response time}.
Usually, this time is are called \emph{pre-movement time}

One possible scheme for the classification of data on evacuation
processes is shown in the following Fig.~\ref{fig:data-evacuation-processes}.
\begin{figure}[hbt]
    \centering
        \includegraphics[width=0.75\textwidth]{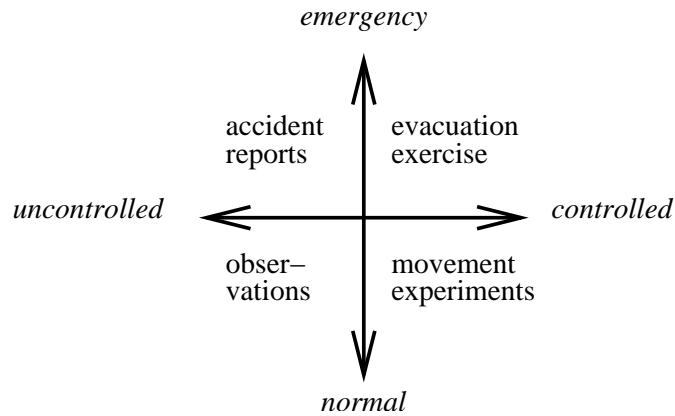}
    \caption{Empirical data can be roughly classified according to 
controlled/uncontrolled and emergency/normal situations.}
    \label{fig:data-evacuation-processes}
\end{figure}
Please note that not only data obtained from uncontrolled or emergency
situations can be used in the context of evacuation assessment.
Especially knowledge about bottleneck capacities (i.e.\ flows through
doors and on stairs) is very important when assessing the layout of a
building with respect to evacuation.
The purpose of empirical data in the context of evacuation processes
(and modelling in general) is threefold
\cite{Galea2004a,ISO-TR-13387-8-19991999}: (1) identify parameters
(factors that influence the evacuation process, e.g.\ bottleneck widths
and capacities), (2) quantify
(calibrate) those parameters, e.g.\ flow through a bottleneck in persons
per meter and second, and (3) validate simulation results, e.g.\ compare
the overall evacuation time measured in an evacuation with simulation
or calculation results.
The validation part is ususally based on data for the evacuation of
complete buildings, aircraft, trains or ships. These are available
from two different sources: (1) full scale evacuation trials and (2)
real evacuations.  Evacuation trials are usually observed and
videotaped. Reports of real evacuation processes are obtained from
eye-witness records and a posteriori incident investigations.
Since the setting of a complete evacuation is not experimental, it is
hardly possible to measure microscopic features of the crowd motion.
Therefore, the calibration of parameters is usually not the main
purpose in evacuation trials, but they are carried out to gain
knowledge about the overall evacuation process and the behavior of the
persons and to identify the governing influences/parameters and
validate simulation results. 

One major concern in evacuation exercises is the well-being of the
participants.  Due to practical, financial, and ethical constraints,
an evacuation trial cannot be realistic by its very nature. Therefore,
an evacuation exercise does not convey the increased stress of a real
evacuation. To draw conclusions on the evacuation process, the walking
speed observed in an exercise should not be assumed to be higher in a
real evacuation \cite{Pauls1995}. Along the same lines of argument, a
simplified evacuation analysis based on e.g. a hydro-dynamic model rather
predicts an evacuation exercise and the same constraints apply for
its results concerning the prediction of evacuation times and the
evacuation process. If the population parameters (like gender, age,
walking speed, etc.) are explicitly stated in the model, increased
stress can be simulated by adapting these parameters.

In summary, evacuation excercises are just too expensive, time consuming,
and dangerous to be a standard measure for evacuation analysis. An
evacuation exercise organized by the UK Marine Coastguard Agency on the
Ro-Ro ferry ``Stena Invicta'' held in Dover Harbor in 1996 cost more
than 10,000 GBP \cite{MCA1997}. This one major argument for the
use of evacuation simulations (resp.\ evacuation analysis based on
hydro-dynamic models and calculations).

\subsubsection{Panic, Herding, and Similar Conjectured Collective Phenomena}

As already mentioned earlier in Sec.~\ref{sec-collective}, the concept
of ``panic'' and its relevance for crowd disasters is rather controversial.  
It is usually used to describe irrational and unsocial behavior.  In the 
context of evacuations empirical evidence shows that this type of behavior 
is rare \cite{Sime1980_tk,Keating1982,Clarke2002,ASA2002}.  
On the other hand there are indications that fear might be 
``contageous'' \cite{Gelder2004}. Related concepts 
like ``herding'' and ``stampede'' seem to indicate a certain similarity
of the behaviour of human crowds
with animal behavior. This terminology is quite often used in the
public media. \emph{Herding} has been described in animal experiments
\cite{Saloma2006} and is difficult to measure in human crowds.
However, it seems to be natural that herding exists in certain
situations, e.g.\ limited visibility due to failing lights or strong
smoke when exits are hard to find.

\textbf{Panic} --- 
As stated earlier, ``panic'' behaviour is usually characterized
by selfish and anti-social behaviour which through contagion affects 
large groups and even leads to completely irrational actions.
Often it is assumed, especially in the media, to occur in situations 
where people compete for scarce or dwindling resources, which in the
case of emergengies are safe space or access to an exit.  
However, this point of view does not stand close scrutiny and it has turned
out that this behaviour has played no role at all in many tragic events
\cite{Keating1982,Johnson1987_tk}.  For these incidents
\emph{crowd disaster} is then a much more appropriate characterisation. 

Furthermore, lack of social behavior seems to be more frequent
during so called ``acquisitive panics'' or ``crazes''
\cite{Smelser1962} than during ``flight panics''. I.e.\ social
behavior seems to be less stable if there is something to gain than if
there is some external danger which threatens all members of a group.
Examples for crazes 
(acquisitive panics) include the Victoria Hall Disaster (1883)
\cite{als_PRED78ENG}, the crowning ceremony of Tsar Nicholas II (1896)
\cite{Schelajew2000_tk}, a governmental Christmas celebration in
Aracaju (2001), the distribution of free Saris in Uttar Pradesh
(2004), and the opening of an IKEA store in Jeddah (2004).  
Crowd accidents which occur at rock concerts and religious events as well 
bear more similarities with crazes than with panics.

However, it is not the case that altruism and cooperation increase with danger.
The events during the capsizing of the MV Estonia 
(see sec. 16.6 of \cite{Laur1997_tk})  show some behavioral threshold:
immediately faced with life-threatening danger, most people struggle for 
their own survival or that of close relatives.

\textbf{Herding} --- Herding in a broad context means ``go with the
flow'' or ``follow the crowd''. Like ``panic'', the term ``herding'' is often
used in the context of stock market crashes, i.e. causing an avalanche
effect. Like ``panic'' the term is usually not well defined and used in
an allegoric way. Therefore, it is advisable to avoid the term in a
scientific context (apart from zoology, of course).  Furthermore,
``herding'', ``stampede'', and ``panic'' have a strong connotation of
``deindividuation''. 
The conjecture of an automatic deindividuation caused by large crowds 
\cite{LeBon1895} has been replaced by a social attachment theory 
(``the typical response to a variety of threats and disasters
is not to flee but to seek the proximity of familiar persons
and places'') \cite{Mawson2005}.

\textbf{Stampede} --- Stampede is -- like herding -- a term from
zoology where herds of large mammals like buffalos collectively run in
one direction and might overrun any obstacles. This is dangerous for
human observers if they cannot get out of the way. The term ``stampede''
is sometimes used for crowd accidents \cite{Johnson1987_tk}, too. It is
furthermore assumed to be highly correlated with panic. When arguing
along those lines, a stampede might be the result of ``crowd panic'' or
vice versa.

\textbf{Shock or density waves} ---
Shock waves are reported for rock concerts \cite{Still2001} and 
religious events \cite{als_helbing07a,AlGadhi2002}. 
They might result in people standing close to each other falling down.
Pressures in dense crowds of up to $4,450~$N/m$^2$ have been reported.

Although empirical data on crowd disasters exist, e.g.\ in the form
of reports from survivors or even video footage, it is almost impossible
to derive quantitative results from them.
Models that aim at describing such scenarios make predictions for
certain counter-intuitive phenomena that should occur.
In the faster-is-slower effect \cite{Helbing2000} a higher
desired velocity leads to a slower movement of a large crowd. In the
freezing-by-heating effect \cite{Helbing2000a} increasing the fluctuations
can lead to a more ordered state. For a thorough discussion we refer
to \cite{Helbing2002,Helbing2000} and references therein.
However, from a statistical point of view there is no sufficient 
data to decide the relevance of these effects in real
emergency situations, also because it is almost impossible to perform
``realistic'' experiments.

\subsubsection{Sources of Empirical Data on Evacuation Processes}

The evacuation of a building can either be an isolated process (due to
fire restricted to this building, a bomb threat, etc.) or it can be
part of the evacuation of a complete area. We will focus on the single
building evacuation, here. For the evacuation of complete areas, e.g.
because of flooding or hurricanes, cf. \cite{Revi2006} \emph{and
  references therein}.

For passenger ships, a distinction between High Speed Craft (HSC),
Ro-Ro passenger ferries, and other passenger vessels (cruise ships) is
made. High Speed Craft do not have cabins and the seating arrangement
is similar to aircraft. Therefore, there is a separate guideline for
HSC \cite{MSC-Circ.11662005}. An performance-based evacuation analysis
at an early stage of design is required for HSC and Ro-Pax. There is
currently no such requirement for cruiseships. For an overview over
IMO's requirements and the historical development up to 2001
cf.~\cite{Dogliani2001}. In addition to the five components for the
overall evacuation time listed above, there are three more specific
for ships: (6) preparation time (for the life-saving appliances, i.e.
lifeboats, life-rafts, davits, chutes), (7) embarkation time, and (8)
launching time. Therefore, the evacuation procedure on ships is more
complex than for buildings. Additionally, SAR (Search And Rescue)
is an integral part of ship evacuation.

For High Speed Craft, the time limit is 17 minutes for evacuation
\cite{HSC-Code2000}, for Ro-Ro passenger ships it is 60 minutes
\cite{MSC-Circ.10332002}, and for all other passenger ships (e.g.
cruise ships) it is 60 minutes if the number of main vertical zones is
less or equal than five and 80 minutes otherwise
\cite{MSC-Circ.10332002}.  For HSC, no distinction is made between
assembly and embarkation phase. 

For aircraft, the approach can be compared to that of HSC. Firstly, an
evacuation test is mandatory and there is a time limit of 90 seconds
that has to be complied to in the test \cite{FAA1990}.

In many countries there is no strict criterion for the maximum evacuation 
time of buildings of buildings. The requirements are based on minimum exit 
widths and maximum escape path lengths.

A number of real evacuations has been investigated and reports are 
publicly available. Among the most recent ones are:  
Beverly Hills Club \cite{Bryan1995}, 
MGM Grand Hotel, \cite{Bryan1995},
retail store \cite{Ashe1999},
department store \cite{Abe1986},
World Trade Center \cite{Grosshandler2004} and \texttt{www.wtc.nist.gov},
high-rise buildings \cite{Pauls1971_cr,Seeger1978_cr},
theatre \cite{Weckman1999} for buildings,
High Speed Craft ``Sleipner'' \cite{NMJP2000} for HSC,
an overview up to 1998 \cite{Owen1998},
exit width variation \cite{Muir1996},
double deck aircraft \cite{Jungermann2000a},
another overview from 2002 \cite{Muir2002} for aircraft, and
for trains \cite{Galea2004a,Schneider2006_cr}.


\section{Modelling}
\label{sec_modelling}

\begin{figure}[h]
  \begin{center}
\includegraphics[width=0.95\textwidth]{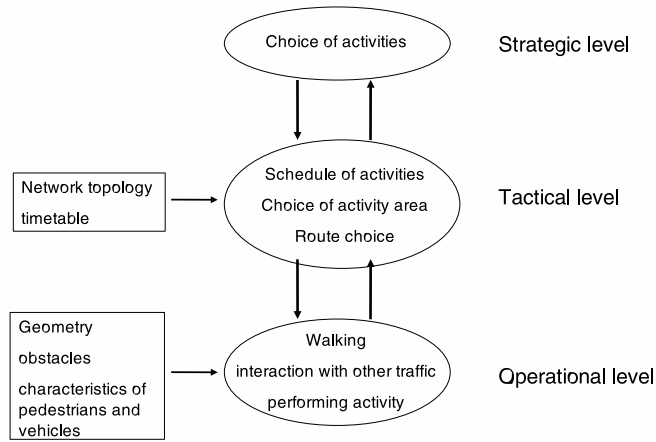}
    \caption{The different levels of modelling pedestrian behaviour
    (after \cite{Hoogendoorn,Daamen2004}).}
\label{fig-ped-levels}
  \end{center}
\end{figure}

A comprehensive theory of pedestrian dynamics has to take into
account three different levels of behaviour
(Fig.~\ref{fig-ped-levels}). At the {\em strategic level},
pedestrians decide which activities they like to perform and the
order of these activities. With the choices made at the strategic
level, the {\em tactical level} concerns the short-term decisions
made by the pedestrians, e.g.\ choosing the precise route taking
into account obstacles, density of pedestrians etc. Finally, the
{\em operational level} describes the actual walking behaviour of
pedestrians, e.g.\ their immediate decisions necessary to avoid
collisions etc.
The processes at the strategic and tactical level are usually
considered to be exogenous to the pedestrian simulation. Here
information from other disciplines (sociology, psychology etc.) is required.
In the following we will mostly be concerned with
the operational level, although some of the models that we are
going to describe allow to take into account certain elements
of the behaviour at the tactical level as well.

Modelling on the operational level is usually based on variations of
models from physics. Indeed the motion of pedestrian crowds has
certain similarities with fluids or the flow of granular materials.
The goal is to find models which are as simple as possible, but
at the same time can reproduce ``realistic'' behaviour in the sense
that the empirical observations are reproduced. Therefore, based
on the experience from physics, pedestrians are often modelled as
simple ``particles'' that interact with each other.

There are several characteristics which can be used to classify
the modelling approaches:
\begin{itemize}
\item {\bf microscopic vs.\ macroscopic:}
In microscopic models each individual is represented separately.
Such an approach allows to introduce different types of pedestrians
with individual properties as well as issues like route choice.
In contrast, in macroscopic models different individuals can not
be distinguished. Instead the state of the system is described
by densities, usually a mass density derived from the positions of
the persons and a corresponding locally averaged velocity. 

\item {\bf discrete vs.\ continuous:}
Each of the three basic variables for a description of a system of
pedestrians, namely space, time and state variable (e.g.\ velocities), 
can be either discrete (i.e.\ an integer number) or continuous (i.e.\ a 
real number). Here all combinations are possible. In a cellular automaton
approach all variables are by definition discrete whereas in hydrodynamic
models all are continuous. These are the most common choices, but other
combinations are used as well. Sometimes for a cellular automata approach also 
a continuous time variable is allowed. In computer simulation this
is realized through a {\em random-sequential update} where at each step
the  particle or site to be updated (moved) is chosen randomly (from
{\em all} particles or sites, respectively). A discrete time is
usually realized through the {\em parallel} or {\em synchronous update} where
all particles or sites are moved at the same time. This introduces
a timescale. In so-called coupled map lattices
time is discrete whereas space and state variables are continuous.

\item {\bf deterministic vs.\ stochastic:}
The dynamics of pedestrians can either be deterministic or stochastic.
In the first case the behaviour at a certain time is completely
determined by the present state. In stochastic models, the behaviour 
is controlled by certain probabilities such that the agents can react 
differently in the same situation. This is one of the lessons learnt
from the theory of complex systems where it has been shown for many
examples that through introduction of stochasticity into rather simple
systems very complex behaviour can be generated. On the other hand,
the stochasticity in the models reflects our lack of knowledge of
the underlying physical processes that e.g.\ determine the decision-making
of the pedestrians. Through stochastic behavioural rules it becomes then
often possible to generate a rather realistic representation of
complex systems like pedestrian crowds.

This ``intrinsic'' stochasticity should be distinguished from ``noise''.
Sometimes external noise terms are added to the {\em macroscopic}
observables, like the position or velocity. Often the main effect of these
terms is to avoid certain special configurations which are considered
to be unrealistic, like completely blocked states. Otherwise the
behaviour is very similar to the deterministic case. For true
stochasticity, on the other hand, the deterministic limit usually
has very different properties from the generic case.

\item {\bf rule-based vs.\ force-based}:
Interactions between the agents can be implemented in at least two
different ways: In a rule-based approach agents make ``decisions'' based
on their current situation and that in their neighbourhood as well as
their goals etc. It focusses on the {\em intrinsic properties} of
the agents and thus the rules are often justified from psychology. 
In force-based models, agents ``feel'' a force exerted by others 
and the infrastructure. They therefore emphasize the {\em extrinsic
properties} and their relevance for the motion of the agents.
It is an physical approach based on the
observation that the presence of others leads to deviations from
a straight motion. In analogy to Newtonian mechanics a force is
made responsible for these accelerations.

Cellular automata are typically rule-based models, whereas e.g.\ the
social-force model belongs to the force-based approaches. However,
sometimes a clear distinction can not be made and many models combine
aspects of both approaches.

\item {\bf high vs.\ low fidelity:}
  {\em Fidelity} here refers to the apparent realism of the modelling
  approach. High fidelity models try to capture the complexity of
  decision making, actions etc.\ that constitute pedestrian motion in
  a realistic way.  In contrast, in the simplest models pedestrians
  are represented by particles without any intelligence. Usually the
  behaviour of these particles is determined by ``forces''. This
  approach can be extended e.g.\ by allowing different ``internal''
  states of the particles so that they react differently to the same
  force depending on the internal state. This can be interpreted as
  some kind of ``intelligence'' and leads to more complex approaches,
  like multi-agent models. Roughly speaking, the number of parameters
  in a model is a good measure for fidelity in the sense introduced
  here, but note that higher fidelity does not necessarily mean that
  empirical observations are reproduced better!
\end{itemize}

It should be mentioned that a clear classification according to
the characteristics outlined here is not always possible. 
In the following we will describe some model classes in more
detail.

\subsection{Fluid-dynamic and gaskinetic models}
\label{sec-ped-models1}

Pedestrian dynamics has some obvious similarities with fluids. E.g the
motion around obstacles appears to follow ``streamlines''. Motion at
intermediate densities is restricted (short-ranged correlations).
Therefore it is not surprising that, very much like for vehicular
dynamics, the earliest models of pedestrian dynamics took inspiration
from hydrodynamics or gas-kinetic theory
\cite{Henderson1974,Helbing1992,Hughes2000,Hughes2002}. Typically
these macroscopic models are deterministic, force-based and
of low fidelity.

Henderson \cite{Henderson1971a,Henderson1974} has tried to establish
an analogy of large crowds with a classical gas. From measurements of
motion in different crowds in a low density (``gaseous'') phase he
found a good agreement of the velocity distribution functions with
Maxwell-Boltzmann distribution \cite{Henderson1971a}.

Motivated by this observation, he has later developed a fluid-dynamic
theory of pedestrian flow \cite{Henderson1974}. Describing the
interactions between the pedestrians as a collision process where the
particles exchange momenta and energy, a homogeneous crowd can be
described by the well-known kinetic theory of gases. However, the
interpretation of the quantities is not entirely clear, e.g.\ what
the analogues of pressure and temperature are in the context of
pedestrian motion. Temperature  could be identified with the
velocity variance, which is related to the distribution of desired 
velocities, whereas the pressure expresses the  desire to move 
against a force in a certain direction.

The applicability of classical hydrodynamical models is based on 
several conservation laws. The conservation of mass, corresponding
to conservation of the total number of pedestrians, 
is expressed through a continuity equation of the form
\begin{equation}
\frac{\partial \rho(\mathbf{r},t)}{\partial t}
+ \nabla\cdot \mathbf{J}(\mathbf{r},t) = 0\, ,
\end{equation}
which connects the local density $\rho(\mathbf{r},t)$ with the
current $\mathbf{J}(\mathbf{r},t)$. This equation can be generalized
to include source and sink terms.
However, the assumption of conservation of energy and momentum is not true 
for interactions between pedestrians which in general do not even satisfy 
Newton's Third Law (``actio $=$ reactio''). In \cite{Helbing1992} several
other differences to normal fluids were pointed out, e.g. the anisotropy of
interactions or the fact that pedestrians usually have an individual
preferred direction of motion.

In \cite{Helbing1992} a better founded fluid-dynamical description
was derived on the basis of a gaskinetic model which describes
the system in terms of a density function $f(\mbfr,\mbfv,t)$.
The dynamics of this function is determined by Boltzmann's 
transport equation that desribes its change for a given state
as difference of inflow and outflow due to binary collisions.

An important new aspect in pedestrian dynamics is the existence of
desired directions of motion
which allows to distinguish different groups $\mu$ of particles. 
The corresponding densities $f_\mu$ change in time due to four 
different effects:
\begin{enumerate}
\item A relaxation term with characteristic time $\tau$
describes tendency of pedestrians to approach their intended velocities.

\item The interaction between pedestrians is modeled by a
  Stosszahlansatz as in the Boltzmann equation. Here pair interactions
  between types $\mu$ and $\nu$ occur with a total rate that is
  proportional to the densities $f_\mu$ and $f_\nu$.

\item Pedestrians are allowed to change from type $\mu$ to $\nu$ 
which e.g.\ accounts for turning left or right at a crossing.

\item Additional gain and loss terms allow to model entrances and exits
where pedestrian can enter or leave the system.
\end{enumerate}

The resulting fluid-dynamic equations derived from this gaskinetic
approach are similar to that of ordinary fluids. However, due to
the different types of pedestrians, corresponding to individuals
who have approximately the same desired velocity, one actually 
obtains a set of coupled equations describing several interacting
fluids. These equations contain additional characteristic terms
describing the approach to the intended velocity and the change
of fluid-type due to interactions in avoidance manoevers.

Equilibrium is approached through the tendency to walk with the
intended velocity, not through interactions as in ordinary fluids.
Momentum and energy are not conserved in pedestrian motion, but the
relaxation towards the intended velocity describes a tendency to
restore these quantities.

Unsurprisingly for a macroscopic approach, the gas-kinetic models
have problems at low densities. For a discussion, see e.g.\
\cite{Helbing1992}.


\subsubsection{Handcalculation method}

For practical applications effective engineering tools have been
developed from the hydrodynamical description. In engineering these
are often called {\em handcalculation methods}. One could also
classify some of them as queing models since the central idea is to
describe pedestrian dynamics as flow on a network with links of
limited capacities.  These methods allow to calculate evacuation times
in a relatively simple way that does not require any simulations.
Parameters entering in the calulations can be adapted to the situation
that is studied. Often they are based on empirical results, e.g.\
evacuation trials.
Details about this kind of models can be found in Sec.~\ref{sec-calcEvT}.

\subsection{Social-Force Models}
\label{sec-sfm}

The social-force model \cite{Helbing1995} is a deterministic
continuum model in which the interactions between pedestrians
are implemented by using the concept of a {\em social force} or
{\em social field} \cite{Lewin51A}. It is based on the idea that
changes in behaviour can be understood in terms of fields or forces.
Applied to pedestrian dynamics the social force $\mathbf{F}^{\rm (soc)}_j$ 
represents the influence of the environment (other pedestrians,
infrastructure) and changes the velocity
$\mathbf{v}_{j}$ of pedestrian $j$. Thus it is responsible for
acceleration which justifies the interpretation as a force.
The basic equation of motion for a pedestrian of mass $m_j$ is then of 
the general form 
\begin{equation}
\frac{d\mbfv_j}{dt} =  \mbff^{\rm (pers)}_j +
\mbff^{\rm (soc)}_j + \mbff^{\rm (phys)}_{j}
\label{eq-sfm-general}
\end{equation}
where $\mbff^{\rm (soc)}_j=\frac{1}{m_j}\mathbf{F}^{\rm (soc)}_j
=\sum_{l\neq j}\mbff^{\rm (soc)}_{jl}$ is the total (specific)
force due to the other pedestrians.
$\mbff^{\rm (pers)}_j$ denotes a ``personal'' force which 
makes the pedestrians attempt to move with their own preferred
velocity $\mbfv^{(0)}_j$ and thus acts as a driving term. 
It is given given by
\begin{equation}
\mathbf{f}^{\rm (pers)}_j = \frac{\mbfv^{(0)}_j-\mbfv_j}{\tau_j}
\end{equation}
where $\tau_j$ reaction or acceleration time.
In high density situations also physical forces
$\mathbf{f}^{\rm (phys)}_{jl}$
become important, e.g.\ friction and compression when pedestrians 
make contact.

The most important contribution to the social force $\mbff^{\rm (soc)}_j$
comes from the territorial effect, i.e.\ the private sphere. Pedestrians
feel uncomfortable if they get too close to others, which
effectively leads to a repulsive force between them. Similar
effects are observed for the environment, e.g.\ people prefer
not to walk too close to walls.

Since social forces are difficult to determine empirically, some
assumptions have to be made. Usually an exponential form is
assumed. Describing the pedestrians as disks of radius $R_j$
and position (of the center of mass) $\mbfr_j$,
the typical structure of the force between the
pedestrians is described by \cite{Helbing2000}
\begin{equation}
\mathbf{f}^{\rm (soc)}_{jl} = A_j \exp\left[
        \frac{R_{jl}-\Delta r_{jl}}{\xi_j}\right]
        \mathbf{n}_{jl}
\label{eq-ped-sfm1}
\end{equation}
with  $R_{jl}=R_j+R_l$, the sum of the disk radia,
$\Delta r_{jl}=|\mbfr_j-\mbfr_l|$,  the distance between the
centers of mass, $\mathbf{n}_{jl}=\frac{\mbfr_j-\mbfr_l
}{\Delta r_{jl}}$, the normalized vector pointing form pedestrian $l$
to $j$.
$A_j$ can be interpreted as strength, $\xi_j$ as the range of 
the interactions. 

The appeal of the social-force model is given mainly by the analogy to 
Newtonian dynamics. For the solution of the equations of motion of 
Newtonian many-particle systems the well-founded molecular dynamics 
technique exists. However, in most studies so far the distinctions 
between pedestrian and Newtonian dynamics are not discussed in detail.
A straightforward implementation of the equations of motion neglecting 
these distinctions can lead to unrealistic movement of single 
pedestrians. For example negative velocities in the main moving 
direction can not be excluded in general even if asymmetric interactions 
(violating Newton's Third Law)
between the pedestrians are chosen. Another effect is the occurrence of 
velocities higher then the preferred velocity $v_j^{(0)}$ due to the 
forces on pedestrians in the moving direction. To prevent this effect 
additional restrictions for the degrees of freedom have to been 
introduced, see for example \cite{Helbing1995}, or the superposition of 
forces has to be discarded \cite{Seyfried2006c}. A general discussion 
of the limited analogy between Newton dynamics and the social-force 
model as well as the consequences for model implementations is still 
missing.


Apart from the ad hoc introduction of interactions the structure
of the social-force model can also be derived from an extremal 
principle \cite{Hoogendoorn03AS,Hoogendoorn2003}. It follows under
the assumption that pedestrian behaviour is determined by the
desire to minimize a certain cost function which takes into account
not only kinematic aspects and walking comfort, but also deviations
from a planned route.

\subsection{Cellular Automata}

Cellular automata (CA) are rule-based dynamical models that are
discrete in space, time and state variable which in the case of
traffic usually corresponds to the velocity. The discreteness in time
means that the positions of the agents are updated in well defined
steps. In computer simulations this is realized through a {\em
  parallel} or {\em synchronous} update where all pedestrians move at
the same time.  The timestep corresponds to a natural timescale
$\Delta t$ which could e.g.\ be identified with some reaction time.
This can be used for the calibration of the model which is essential
for making quantitative predections.  A natural space discretization
can be derived from the maximal densities observed in dense crowds
which gives the minimal space requirement of one person. Usually each
cell in the CA can only be occupied by one particle (exclusion
principle)  so that this space requirement can be identified with the
cell size. In this way, a maximal density of 6.25~P/m$^2$
\cite{Weidmann1993} leads to a cell size of $40\times 40$~cm$^2$. 
Sometimes finer discretizations are more appropriate (see 
Sec.~\ref{sec_theor-result}). In this case pedestrians correspond to 
extended particles that occupy more than one cell (e.g.\ four cells).
The exclusion principle 
and the modelling of humans as non-compressible particles mimicks 
short-range repulsive interactions, i.e.\ the ``private-sphere''.

\begin{figure}[htb]
  \begin{center}
    \includegraphics[width=0.55\textwidth]{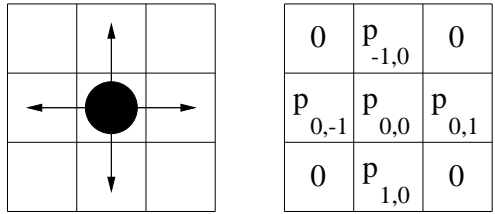}
    \caption{A particle, its possible directions of motion and the
corresponding transition probabilities $p_{ij}$ for the case of a 
von Neumann neighbourhood.
} \label{fig-ped-transition}
  \end{center}
\end{figure}

The dynamics is usually defined by rules which specify the transition 
probabilities for the motion to one of the neighbouring cells 
(Fig.~\ref{fig-ped-transition}). The models differ in the specification 
of these probabilites as well in that of the ``neighbourhood''. For
deterministic models all except of one probability are zero.



The first cellular automata (CA) models
\cite{FukuiI99A1,Muramatsu1999,KluepfelMKWS00A,Blue2000a} for
pedestrian dynamics can be considered two-dimensional variants of the
asymmetric simple exclusion process (ASEP) (for reviews, see
\cite{Derrida98A,Schuetz01A,BlytheE07A}) or models for city or
highway traffic \cite{NagelS92A,BihamML92A,Chowdhury2000} based on it.
Most of these models represent pedestrians by particles withouth
any internal degrees of freedom. They can move to one of the
neighbouring cells based on certain transition probabilities which
are determined by three factors: (1) the desired direction of motion,
e.g.\ to find the shortest connection, (2) interactions with other
pedestrians, and (3) interactions with the infrastructure (walls,
doors, etc.).


\subsubsection{Fukui-Ishibashi model}

One of the first CA models for pedestrian dynamics has been proposed by 
Fukui and Ishibashi \cite{FukuiI99A1,FukuiI99A2} 
and is based on a two-dimensional variant of the ASEP. 
They have studied bidirectional motion in a long corridor where
particles moving in opposite directions are updated alternatingly.
Particles move deterministically in their desired direction, only if the 
desired cell is occupied by an oppositely moving particle
they make a random sidestep.

Various extensions and variations of the model have bee proposed,
e.g.\ an asymmetric variant \cite{Muramatsu1999} where walkers prefer 
lane changes to the right,
different update types \cite{Weifeng2003}, simultaneous (exchange) 
motion of pedestrians standing ``face-to-face'' \cite{JianLD05A},
or the possibility of backstepping \cite{Maniccam05A}. The influence of
the shape of the particles has been investigated in \cite{NagaiN06A}.
Also other geometries \cite{Muramatsu2000e,Tajima2002}
and extensions to full 2-dimensional motion have been studied in
various modifications \cite{Muramatsu2000d,Maniccam03A,Maniccam05A}


\subsubsection{Blue-Adler model}

The model of Blue and Adler \cite{Blue2000a,BlueA02A} is based on a
variant of 
the Nagel-Schreckenberg model \cite{NagelS92A} of highway traffic.
Pedestrian motion is considered in analogy to a multi-lane highway.  
The structure of the rules is similar to the basic two-lane rules
suggested in \cite{RickertNSL96A}. The update is performed in four
steps which are applied to all pedestrians in parallel. In the first
step each pedestrian chooses a preferred lane. In the second step the
lane changes are performed. In the third step the velocities are
determined based on the available gap in the new lanes. Finally, in
the fourth step the pedestrians move forward according to the
velocities determined in the previous step.

In counterflow situations head-on-conflicts occur. These are resolved
stochastically and with some probability opposing pedestrians are
allowed to exchange positions within one timestep. Note that the
motion of a single pedestrian (not interacting with others) is
deterministic otherwise.

Different from the Fukui-Ishibashi model motion is not restricted
to nearest-neighbour sites. Instead pedestrians can have different
velocities $v_{\rm max}$ which correspond to the maximal number of
cells they are allowed to move forward. In contrast to vehicular
traffic, acceleration to  $v_{\rm max}$ can be assumed to be
instantaneous in pedestrian motion.

In order to study the effects of inhomogeneities, the pedestrians
are assigned different maximal velocities $v_{\rm max}$. Fast walkers
have $v_{\rm max}=4$, standard walkers $v_{\rm max}=3$ and slow
walkers $v_{\rm max}=2$. The cell size is assumed to be
50~cm$\times$~50~cm. The best agreement with empirical observations
has been achieved with 5\% slow and 5\% fast walkers \cite{BlueA02A}.
Furthermore the fundamental diagram in more complex situations,
like bi- or four-directional flows have been investigated.


\subsubsection{Gipps-Marksj\"os model}

A more sophisticated discrete model has been suggested by Gipps and
Marksj\"os \cite{Gipps1985} already in 1985. One motivation
for developing a discrete model was the limited computer power
at that time. Therefore a discrete model, which reproduces
the properties of pedestrian motion realistically, was in many
respects a real improvement over the existing continuum approaches.

Interactions between pedestrians are assumed to be repulsive anticipating
the idea of social forces (see Sec.~\ref{sec-sfm}).
The pedestrians move on a grid of rectangular cells of size $0.5\times
0.5$~m. To each cell a score is assigned based on its proximity
to other pedestrians. This score represents the repulsive interactions
and the actual motion is then determined by the competition between
these repulsion and the gain of approaching the destination.
Applying this procedure to all pedestrians, to each cell a potential 
value is assigned which is the sum of the individual contributions.
The pedestrian then selects the cell of its nine neighbours (Moore
neighbourhood) which leads to the maximum benefit. This benefit is
defined as the difference between the gain of moving closer to
the destination and the cost of moving closer to other pedestrians
as represented by the potential. This requires a suitable chosen
gain function $P$. 

The updating is done sequentially to avoid conflicts of
several pedestrians trying to move to the same position. In order
to model different velocities, faster pedestrians are updated more
frequently. Note that the model dynamics is deterministic.


\subsubsection{Floor field CA}\label{sec-ped-CA2}

The floor field CA 
\cite{Burstedde2001a,Schadschneider2002,Burstedde2002,Kirchner2002b} 
can also be considered as an extension of the ASEP. However, the
transition transition probabilities to neighbouring cells are no longer
fixed but vary dynamically.
This is motivated by the process of chemotaxis 
(see \cite{Benjacob97A} for a review) used by some insects (e.g.\ ants)
for communication. They create a chemical trace to guide 
other individuals to food sources. In this way a complex trail system
is formed that has many similarites with human transport networks.

In the approach of \cite{Burstedde2001a} the pedestrians also create a
trace. In contrast to chemotaxis, however, this trace is only virtual
although one could assume that it corresponds to some abstract
representation of the path in the mind of the pedestrians. Although
this is mainly a technical trick which reduces interactions to local
ones that allow efficient simulations in arbitrary geometries, one
could also think of the trail as reprentation of the paths in the mind
of a pedestrian.  The locality becomes important in complex geometries
as no algorithm is required to check whether the interaction between
particles is screened by walls etc.  The number of interaction terms
always grows linearly with the number of particles.

\begin{figure}[htb]
  \begin{center}
    \includegraphics[width=0.3\textwidth]{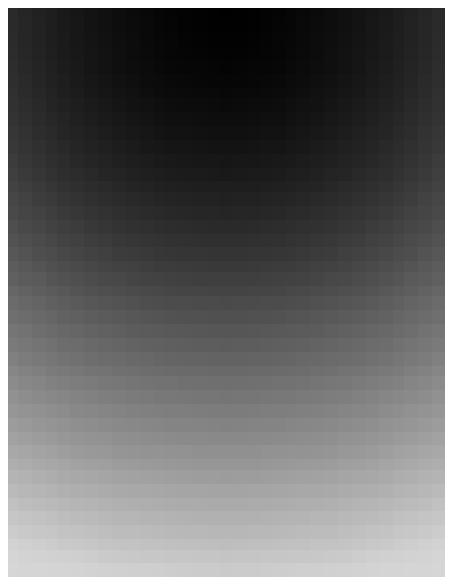}
    \qquad\qquad\qquad
    \includegraphics[width=0.3\textwidth]{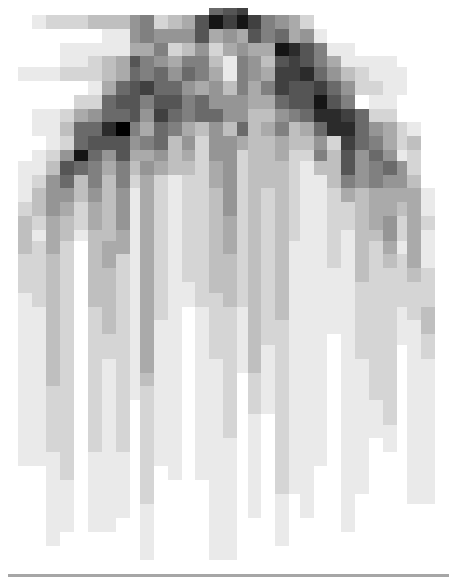}
    \caption{Left: Static floor field for the simulation of an evacuation
from a large room with a single door. The door is located in the
middle of the upper boundary and the field strength is increasing
with increasing intensity. Right: Snapshot of the dynamical floor field 
created by people leaving the room.
}
\label{fig-ped-sff}
  \end{center}
\end{figure}

The translation into local interactions is achieved
by the introduction of so-called {\em floor fields}. The transition
probabilities for all pedestrians depend on the strength of the
floor fields in their neighbourhood in such a way that transitions
in the direction of larger fields are preferred.
The {\em dynamic floor field} $D_{ij}$ corresponds to a virtual trace
which is created by the motion of the pedestrians and in turn
influences the motion of other individuals. Furthermore it has its own
dynamics, namely through diffusion and decay, which leads to a
dilution and finally the vanishing of the trace after some time.  The
{\em static floor field} $S_{ij}$ does not change with time since it
only takes into account the effects of the surroundings. Therefore it
exists even without any pedestrians present. It allows to model e.g.\ 
preferred areas, walls and other obstacles.
Fig.~\ref{fig-ped-sff} shows the static floor field used
for the simulation of evacuations from a room with a single door.  Its
strength decreases with increasing distance from the door. Since the
pedestrian prefer motion into the direction of larger fields, this is
already sufficient to find the door.

Coupling constants control the relative influence of both fields.
For a strong coupling to the static field pedestrians will choose
the shortest path to the exit. This corresponds to a 'normal' situation.
A strong coupling to the dynamic field implies a strong herding
behaviour where pedestrians try to follow the lead of others.
This often happens in emergency situations.

The model uses a fully parallel update. Therefore conflicts can occur 
where different particles choose the same destination cell. 
This is relevant for high density situations and happens in all
models with parallel update if motion in different directions is
allowed. Conflicts have been considered a technical problem for
a long time and usually the dynamics has been modified in order
to avoid them. The simplest method is to update pedestrians sequentially
instead of using a fully parallel dynamics. However, this leads
to other problems, e.g.\ the identification of the relevant 
timescale. Therefore it has been suggested in \cite{KirchnerNNS03A}
to take these conflicts seriously as an important part of the
dynamics.

\begin{figure}[h]
\begin{center}
\includegraphics[width=0.7\textwidth]{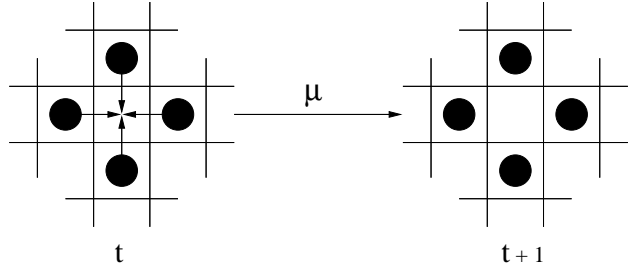}
\end{center}
\vspace{-0.4cm} \caption[]{Refused movement due to the friction
parameter $\mu$ (for $m=4$).} \label{plot_0}
\end{figure}

For the floor field model it has been shown in \cite{Kirchner2003b}
that the behaviour becomes more realistic if not all conflicts are
resolved in the sense that one of pedestrian is allowed to move
whereas the others stay at their positions. Instead with probability
$\mu \in [0,1]$, which is called friction parameter, the movement of
{\em all} involved pedestrians is denied \cite{Kirchner2003b} (see
Fig.~\ref{plot_0}).
This allows to describe clogging effects between the pedestrians in a
much more detailed way \cite{Kirchner2003b}. $\mu$ works as some kind
of local pressure between the pedestrians. If $\mu$ is high, the
pedestrians handicap each other trying to reach their desired target
sites. This local effect can have enormous influence on macroscopic
quantities like flow and evacuation time \cite{Kirchner2003b}. Note
that the kind of friction introduced here only influences interacting
particles, not the average velocity of a freely moving pedestrian.

Surprisingly the qualitative behaviour of the floor field model and 
the social-force model is very similar despite the fact that 
the interactions are very different. In the floor field model
interactions are attractive whereas they are repulsive in the 
social-force model. However, in the latter interactions are 
between particle densities. In contrast in the floor field model
the particle density interacts with the velocity density.

\subsection{Other Approaches}

\subsubsection{Lattice-gas models}\label{sec-ped-models3}

In 1986, Frisch, Hasslacher, and Pomeau \cite{FrischHP86A} have shown that 
one does not have to take into account the detailed molecular motion within 
fluids in order to obtain a realistic picture of (2d) fluid dynamics. 
They proposed a lattice gas model \cite{RothmanZ94A,RothmanZ97A}
on a triangular lattice with hexagonal symmetry which is similar in
spirit to CA models, but the exclusion principle is relaxed: Particles
with different velocities are allowed to occupy the same site. Note
that the allowed velocities differ only in the direction, not the
absolute value. The dynamics is based on a succession of collision and
propagation that can be chosen in such a way that the coarse-grained
averages of this microscopic dynamics is asymptotically equivalent to
the Navier-Stokes equations of incompressible fluids.

In \cite{MarconiC02A} a kind of mesoscopic approach inspired by these
lattice gas models has been suggested as a model for pedestrian
dynamics. In analogy with the description of transport phenomena in
fluids (e.g.\ the Boltzmann equation) the dynamics is based on a
succession of collision and propagation.
\begin{figure}[h]
  \begin{center}
\includegraphics[width=0.96\textwidth]{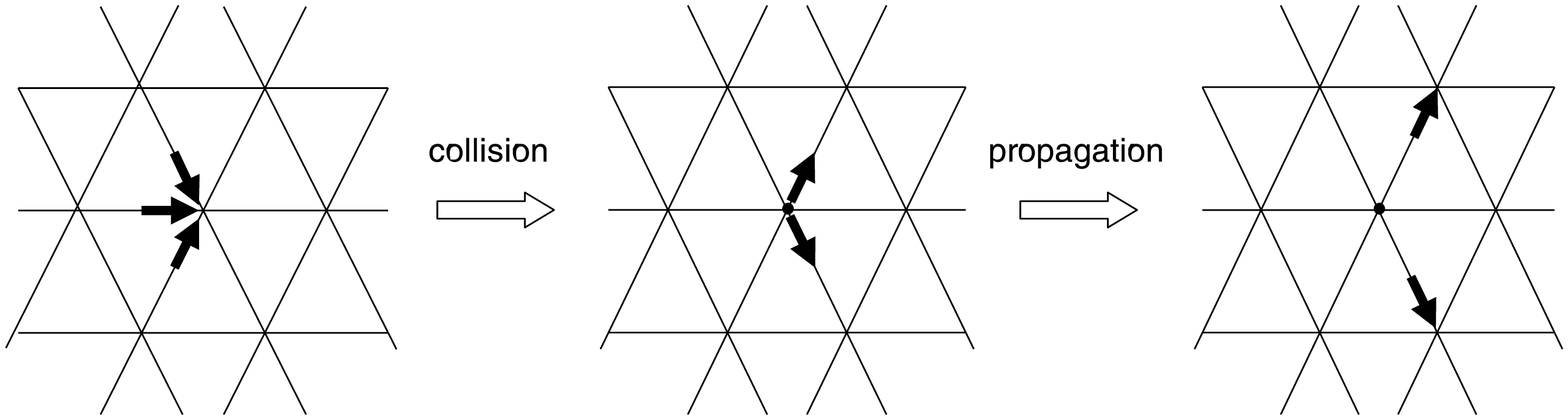}
  \end{center}
    \caption{The dynamics of lattice gas models proceeds in two steps.
    Pedestrians coming from neighbouring sites interact in the collision
    step where velocities are redistributed. In the propagation step
    the pedestrians move to neighbour sites in the directions determined
    in the collision step.}
\label{fig-ped-lattgas}

\end{figure}
Pedestrians are modelled as particles moving on a triangular lattice
which have a preferred direction of motion $\mbfc_F$. However, the
particles do not follow strictly this direction by have also a
tendency to move with the flow. Furthermore at high densities the
crowd motion is influenced by a kind of friction which slows down the
pedestrians. This is achieved by reducing the number of individuals
allowed to move to neighbouring sites.

As in a lattice gas model \cite{RothmanZ97A}, the dynamics now
consists of two steps. In the {\em propagation step} each pedestrian
moves to the neighbour site in the direction of its velocity vector.
In the {\em collision step} the particles interact and new velocities
(directions) are determined. In contrast to physical systems,
momentum etc.\ does not need to be conserved during the collision step.
These considerations lead to a collision step that takes into
account the favorite direction $\mbfc_F$, the local density 
(the number of pedestrians at the collision site), and a quantity 
called mobility 
at all neighbour sites 
which is a normalized measure of the local flow after the collision.


\subsubsection{Optimal-velocity model}\label{sec-ped-models4}

The optimal velocity (OV) model originally introduced for the description
of highway traffic can be generalized to higher dimensions 
\cite{NakayamaHS05A} which allows its application to pedestrian dynamics.

In the two-dimensional extension of the OV model 
the equation of motion for particle $i$ is given by
\begin{eqnarray}
\frac{d^2}{dt^2}{\bf x}_i(t) &=& a \biggl\{{\bf V}_0+\sum_j {\bf
V}({\bf x}_j(t)-{\bf x}_i(t)) - \frac{d}{dt}{\bf x}_i(t)\biggr\}\,,
\label{eq-ped-ovm2d}
\end{eqnarray}
where ${\bf x}_i =(x_i,y_i)$ i the position of particle $i$.  
It can be considered as a special case of the general social-force
model (\ref{eq-sfm-general}) without physical forces.
The optimal-velocity function 
\begin{eqnarray}
&& {\bf V}({\bf x}_j-{\bf x}_i) = f(r_{ij})(1+\cos\varphi)\ {\bf n}_{ij}, \\
&& f(r_{ij}) = \alpha \{\tanh\beta (r_{ij} - b) + c \},
\label{eq-ped-OV-def2}
\end{eqnarray}
where $r_{ij} = |{\bf x}_j-{\bf x}_i|$, $\cos\varphi =
(x_j-x_i)/r_{ij}$ and ${\bf n}_{ij} = ({\bf x}_j-{\bf x}_i)/r_{ij}$
is determined by interactions with other pedestrians.
${\bf V}_0$ is a constant vector that represents a `desired
velocity' at which an isolated pedestrian would move.
The strength of the interaction depends on the distance $r_{ij}$
between the $i$th and $j$th particles, and on the angle $\varphi$
between the directions of ${\bf x}_j-{\bf x}_i$ and the current
velocity $\frac{d}{dt}{\bf x}_i$. Due to the term $(1+\cos\varphi)$,
a particle reacts more sensitively to particles in front than those behind. 

Now two cases can be distinguished, namely repulsive and attractive
interactions. The former is relevant for pedestrian dynamics whereas
the latter is more suitable for biological motion. Therefore for
pedestrian motion one chooses $c=1$ which implies $f< 0$.

A detailed analysis \cite{NakayamaHS05A} shows that the model exhibits
a rich phase diagram including the formation of various patterns.


\subsubsection{Other models}

We briefly mention a few other model approaches that have been
suggested. In \cite{Bolay1998} a disretized version of the social-force
model has been introduced and shown to reproduce qualitatively
the observed collective phenomena.

In \cite{Okazaki1993} a magnetic force model has been proposed
where pedestrians and their goals are treated as magnetic poles of
opposite sign.

Another class of models is based on ideas from queuing theory.
In principle, some handcalculation methods can be considered 
as a macroscopic queuing model. 
Typically rooms are represented as nodes in the queuing 
network and links correspond to doors. In microscopic approaches 
in the movement process each agent chooses a new node, e.g.\ according
to some probability \cite{Lovas1994}.

\subsection{Theoretical Results}
\label{sec_theor-result}

As emphasized in Sec.~\ref{sec-collective}, the collective effects
observed in the motion of pedestrian crowds are a direct consequence of the
microscopic dynamics. These effects are reproduced quite well by some
models, e.g. the social-force and floor-field model, at least on 
a qualitative level. 
As mentioned before, the qualitative behaviour of the two
models is rather similar despite the very different implementation
of the interactions. This indicates a certain robustness of the
collective phenomena observed.

\begin{figure}[h]
  \begin{center}
    \includegraphics[width=0.9\textwidth]{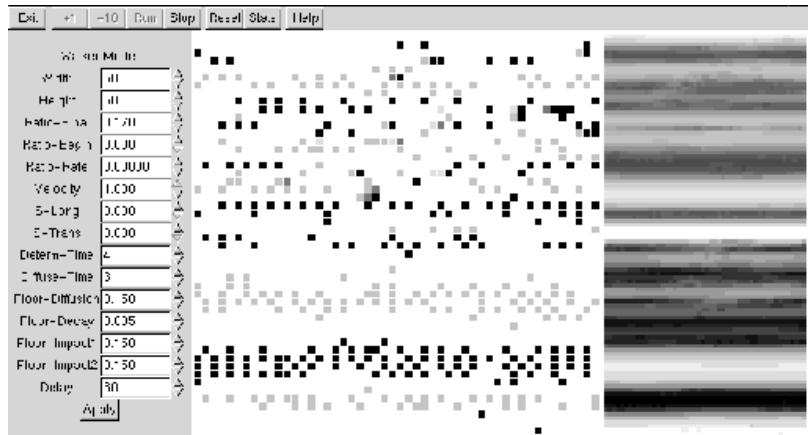}
  \end{center}
    \caption{Lane formation in the floor-field model. The central window
    is the corridor and the light and dark squares are right- and
    left-moving pedestrians, respectively.  In the bottom part well-separated
    lanes can be observed whereas in the top part the motion is still
    disordered.  The right part of the figure shows the
    floor fields for the right-movers (upper half) and left-movers
    (lower half).}
\label{fig-CA-lanes}
\end{figure}

As an example we discuss the formation of lanes in counterflow formation.
Empirically one observes a strong tendency to follow immediately in the 
``wake'' of another person heading into the same direction. 
Such lane formation was reproduced in the social-force model 
\cite{Helbing1995,Helbing2000a} as well as in the floor-field
model \cite{Burstedde2001a,Kaufman2007_tk} (see  Fig.~\ref{fig-CA-lanes}). 
While the formation of lanes in general is essential to avoid deadlocks 
and thus keep the chance to reproduce realistic fluxes, the number of 
direction changes per meter cross section is a parameter which in
reality crucially depends on the situation \cite{Kaufman2007_tk}: The 
longer a counterflow situation is assumed to persist, the less lanes per
meter cross section can be found. The correct reproduction of 
counterflow is an issue for an accomodating animation, but
more or less unimportant for the macroscopic observables. This is
probably the main reason why there seems to have been not much effort
put into the attempt to reproduce different ``kinds'' of lane formation
in a controlled, situation-dependent manner.

On the quantitative side, the fundamental diagram is the first and most
serious test for any model. Since most quantitative results rely on the
fundamental diagram it can be considered the most important characteristics 
of pedestrian dynamics.
It is not only relevant for movement in a corridor or through a 
bottleneck, but also as an important determinant of evacuation times. 
However, as emphasized earlier, there is currently no consensus on 
the empirical form of the fundamental diagram. Therefore a calibration
of the model parameters is currently not sensible.

Most cellular automata models are based on the asymmetric simple
exclusion process. This strictly one-dimensional stochastic process
has a fundamental diagram which is symmetric around density $\rho=1/2$.
Lane changes in two-dimensional extensions lead only to
a small shift towards smaller densities. Despite the discrepancies
in the empirical results, an almost symmetric fundamental diagram
can be excluded.

Based on the experience with modelling of highway traffic 
\cite{NagelS92A,Chowdhury2000} therefore models with higher speeds
have been introduced which naturally lead to an asymmetric
fundamental diagram. Typically this is implemented
by allowing the agents to move more than one cell per update step
\cite{Kluepfel2003,Kirchner2003,Kirchner2004,YamamotoKN06A,YamamotoKN07A,Kretz2007a}.  These model variants have been shown to be flexible enough
to reproduce e.g.\ Weidmann's fundamental diagram for the
flow in a corridor \cite{Weidmann1993} with high precision. 
Usually in the simulations a homogeneous population is assumed.
However in reality different pedestrians have different properties
like walking speed, motivation etc. This is easily taken into account
in every microscopic model.
There are many parameters that could 
potentially have an influence on the fundamental diagram. However, the 
current empirical situation does not allow to decide this question.

Another problem occuring in CA models has its origin in the
discreteness of space. Through the choice of the lattice
discretization space is no longer isotropic.  Motion in directions
not parallel to the main axis of the lattice are difficult to realize
and can only be approximated by a sequence of steps parallel to the
main directions. 

Higher velocities require also the extension of the neighbourhood
of a particle which is no longer identical to the cells adjacent
to the current position. A natural definition of ``neighbourhood''
corresponds to those cells that could be reached within one timestep.
In this way the introduction of higher velocities also reduces
the problem of space isotropy as the neighbourhoods become more
isotropic for larger velocities.

Other solutions to this problem have been proposed. One way is to count
the number of diagonal steps and let the agent suspend from moving
following certain rules which depend on the number of diagonal steps
\cite{SchultzLF06A}.  A similar idea is to sum up the real distance
that an agent has moved during one round: A diagonal step counts
$\sqrt{2}$ and a horizontal or vertical one $1$.  An agent has to
finish its round as soon as this sum is bigger than its
speed \cite{Kluepfel2003}. A third possibility - which works for
arbitrary speeds - is to assign selection probabilities to each of the
four lattice positions which are adjacent to the exact final position
\cite{YamamotoKN06A,YamamotoKN07A}.  Naturally these probabilities are
proportional to the square area between the exact final position and
the lattice point, as in this case the probabilities are normalized by
construction if one has a square lattice with points on all integer
number combinations.  However, one also could think of other methods
to calculate the probability.

For the social-force model, the specification of the repulsive
interaction (with and without hard core, exponential or reciprocal
with distance) as well as the parameter sets for the forces
changes in different publications
\cite{Helbing1995,Helbing2000a,Helbing2000,Moln'ar1995}.  
In \cite{Helbing2002} the authors state that
``most observed self-organization phenomena are quite insensitive to
the specification of interaction forces''. 
However, at least for the fundamental diagram, a relation connected with 
all phenomena in pedestrian dynamics, this statement is questionable. 
As remarked in \cite{Werner03} the reproduction of the fundamental
diagram ``requires a less simple specification of the repulsive 
interaction forces''. 
Indeed in \cite{Seyfried2006c} it was shown that the choice of hard-core 
forces or repulsive soft interactions as well as the particular parameter 
set can strongly influence the resulting fundamental diagram regarding
qualitative as well as quantitative effects.

Also a more realistic behaviour at higher densities requires
a modification of the basic model. Here the use of density-dependent
desired velocities leads to a reduction of the otherwise unrealistically
large number of collisions \cite{Bolay1998}.

The particular specification of forces and the previously mentioned problem
with Newton's Third law can lead in principle to some unwanted effects,
like momentary velocities larger than the preferred velocity
\cite{Helbing1995} or the penetration of pedestrians into each other
or into walls \cite{Lakoba2005}.  It is possible that these effects
can be suppressed for certain parameter sets by contact or friction
forces, but the general appearance is not excluded. Only in the first
publication \cite{Helbing1995} restrictions for the velocity are
explicitly formulated to prevent velocities larger the the intended
speed and other authors tried to improve the model by introducing more
parameters \cite{Lakoba2005}.  But additional parameter and artifical
restrictions of variables diminish the simplicity and thus the
attractiveness of the model.
A general discussion how to deal with these problems of the social-force
model and a verification that the observed phenomena are not limited to 
a certain specification of the interaction and a special
parameter set is up to now still missing.

While the realistic reproduction within the empirical range of these
macroscopic observables, especially the fundamental diagram, is
absolutely essential to guarantee safety standards in evacuation
simulations, and while a user should always be distrustful of models
where no fundamental diagram has ever been published, it is by no
means sufficient to exclusively check for the realism of macroscopic
observables. On the microscopic level there is a large amount of
phenomena which need to be reproduced realistically, be it just to
make a simulation animation look realistically or be it for the reason
that microscopic effects can often easily influence macroscopic
observables.

If one compares simulations of bottleneck flows with real events, one
observes that in simulations the form of the queue in front of
bottlenecks is often a half-circle, while in reality it is drop- or
wedge-shaped. In most cases this discrepancy probably does not have an
influence on the simulated evacuation time, but it is interesting to
note, where it originates from. Most simulation models implicitly or
explicitly use some kind of utility maximation to steer the
pedestrians -- with the utility being aforemost inversely proportional
to the distance from the nearest exit. This obviously leads to
half-circle-shaped queues in front of bottlenecks. So wherever one
observes queues different than half-circles, people have exchanged
their normal ``utility function based on the distance'' with something 
else. One such alternative utility function could be that people are just
curious what is inside or behind the bottleneck, so they need to seek
a position where they can look into it. A more probable explanation
would be that in any case it is the time distance not the spatial
distance which is seeked to be minimized.  As anyone knows about the
inescapable loss in time a bottleneck means for the whole waiting
group, the precise waiting spot is not that important. However, in
societies with a strong feeling for egality, people strongly would
wish to equally distribute the waiting time and keep a
first-in-first-out principle, which can best be accomplished and
controlled when the queue is more or less one-dimensional,
respectively just as wide as the bottleneck itself.

Finally it should be mentioned that theoretical investigations 
based on simulations of models for pedestrian dynamics have lead to the 
prediction of some surprising and counter-intuitive collective phenomena, 
like the reduction of evacuation times through additional columns near exits 
(see Sec.~\ref{BCK}) or the faster-is-slower \cite{Helbing2000}
and freezing-by-heating effect \cite{Helbing2000a}.
However, so far the empirical evidence for the relevance or even
occurance of these effects in real situations is rather scarce.


\section{Applications}

In the following section we discuss more practical aspects of 
based on the modelling concepts presented in Sec.~\ref{sec_modelling}.
Tools of different sophistication have been developed that are 
nowadays routinely used in safety analysis. The latter becomes more
and more relevant since many public facilities have to fulfill certain
legal standards. As an example we mention aircrafts which have to be 
evacuated within 90 seconds. The simulations etc.\ are already used in 
the planning stages because changes of the design at a later stage are
difficult and expensive.

For this kind of safety analysis tools of different sophistication have 
been developed. Some of them mainly are able to predict just evacuation
times whereas others are based on microscopic simulations which allow
also to study various external influences (fire, smoke, ...) in much
detail. 

\subsection{Calculation of Evacuation Times}
\label{sec-calcEvT}

The basic idea of handcalculation methods has already briefly been
described at the end of Sec.~\ref{sec-ped-models1}. Here we want to
discuss its practical aspects in more detail.

The approach has been developed since the middle of the 1950s
\cite{Togawa1955}. The basic idea of these methods is the assumption
that people can be calculated or behave like fluids. Knowledge of the
flow (see Equ. \ref{FLOW_ALS}) and the technical data of the facility are then
sufficient to evaluate evacuation times etc. 

Handcalculation methods can be divided in two major approaches:
methods with ``dynamic'' flow
\cite{Weidmann1993,Fruin1971,Roitman1966_cr,Predtetschenski1971,Predtetschenski1972_cr,Kendik1984_cr,Kendik1986a_cr,Kendik1983_cr,Nelson2002,Galbreath1969_cr}
and methods with ``fixed'' flow
\cite{Mueller1970_cr,NFPA_cr,Melinek1975_cr,Mueller1968_cr,Mueller1966_cr,Mueller1966a_cr,Pauls1995,Seeger1978_cr,Togawa1955}.
As methods with ``dynamic'' flow we call methods where the pedestrian
flow is dependent from the density of the pedestrian stream (see Sec.
\ref{sec-observables}) in the selected facility, thus the flow can be
obtained from fundamental diagrams (see Sec.~\ref{FUNDDIA}) or it is
explicitly prescribed in the chosen method. This flow can change
during movement through the building, e.g. by using stairs, thus the
pedestrian stream has a ``dynamic'' flow. Methods with ``fixed'' flow
do not use this concept of relationship between density and flow. In
this methods selected facilities (e.g. stairs or doors) has a fixed
flow which is independent from the density, that is usually not used
in this methods. The ``fixed'' flow usually based upon empirical and
measured data of flow, which are specified for a special type of
buidling, like high-rise buildings or railway stations, for example.
Because of much simplifications in these ``fixed'' flow methods a
calculation can always be done very fast.

Methods with ``dynamic'' flow allow the user to describe the condition
of the pedestrian flow in every part of the selected building or
environment, because they are mostly based upon the continuity
equation, thus it is possible to calculate different kind of
buildings. This allows the user to calculate transitions from wide to
narrow, floor to door, floor to stair, etc.  The disadvantage is that
some these methods are very elaborate and time-intensive. But not in
general a method with ``dynamic'' flow is complicated to calculate,
thus we want to divide handcalculation methods in simple
\cite{Mueller1970_cr,NFPA_cr,Melinek1975_cr,Roitman1966_cr,Nelson2002,Predtetschenski1972_cr,Fruin1971,Weidmann1993,Pauls1995,Mueller1968_cr,Mueller1966_cr,Mueller1966a_cr,Galbreath1969_cr,Seeger1978_cr,Togawa1955}
and complex
\cite{Predtetschenski1971,Kendik1984_cr,Kendik1986a_cr,Kendik1983_cr}
for evacuation calculation.  All of these handcalculation methods are
able to predict total evacuation times for a selected building, but
differences between different methods are still alive. Thus the user
has to ensure that he is familiar with assumptions made by each method
to ensure that a result is interpreted in a correct way
\cite{Rogsch2007_cr}.

\subsection{Simulation of Evacuation Processes}
\label{sec:evacuation_simulations}

Before we go into the details of evacuation simulation, let us briefly
clarify its scope and limitations and contrast it to other methods
used in evacuation analysis. When analyzing evacuation processes, three
different approaches can be identified:
(1) risk assessment, (2) optimization, and (3) simulation. The aim and
result of risk-assessment is a list of events and their consequences
(e.g. damage, financial loss, loss of life), i.e. usually an event tree
with probabilities and expectation values for financial loss.
Optimization aims at, roughly speaking, minimizing the evacuation time
and reducing the area and duration of congestion. And finally,
simulation describes a system with respect to its function and behavior
by investigating a model of the system. This model is usually
non-analytic, does not provide explicit equations for the calculation
of, e.g. evacuation time.
Of course, simulations are used for ``optimization'' in a more general
sense, too, i.e. they can be part of an optimization. This holds for
risk assessment, too, if simulations are used to determine the outcomes
of the different scenarios in the event tree.

In evacuation analysis the system is, generally speaking, a group of
persons in an environment. More specifically, four components
(sub-systems/sub-models) of the system \emph{evacuation process} can
be identified: (1) geometry, (2) environment, (3) population, and (4)
hazards \cite{Galea2004a}. Any evacuation simulation must at least take
into account (1) and (3).
The behavior of the persons (which can be described on the strategic,
tactical, and operational level --- see Sec.~\ref{sec_modelling})
level is part of the population sub-model.
An alternative way of describing the behavior is according to its
algorithmic representation:
no behavior modeling - functional analogy - implicit representation
(equation) - rule based - artificial intelligence \cite{Galea2004a}.

Hazards are in the context of evacuation first of all fire and smoke,
which then require a toxicity sub-model, e.g. the fractional effective
dose model (FED), to assess their physiological effect of toxic gases
and temperature \cite{DiNenno2002}.
Further hazards to take into account might be earthquakes, floodings, or
in the case of ships, list, heel, or roll motion.
The sub-model environment comprises all other influences that affect the
evacuation process, e.g. exit signs, surface texture, public address
system, etc.

In summary, aims of an evacuation analysis and simulation are to provide
feedback and hints for improvement at an early stage of design,
information for safer and more rigorous regulations, improvement of
emergency preparedness, training of staff, and accident investigation
\cite{Galea2004a}. They usually do not provide direct results on the
probability of a scenario or a systematic search for optimal geometries.

\subsubsection{Calculation of Overall Evacuation Time, Identification 
of Congestion, and Corrective Actions}

The scope of this section is to show general results that can be
obtained by evacuation simulations. They are general in the sense that
they can basically be obtained by any stochastic and microscopic model,
i.e. apart from these two requirements, the results are not model
specific. In detail, five different results 
of evacuation simulations can be distinguished:
(1) distribution of evacuation times, (2) evacuation curve (number of
persons evacuated vs. time), 
(3) sequence of the evacuation (e.g. snapshots/screenshots at specific
times, e.g. every minute), and (4) identification of congestion,
usually based on density and time. Especially the last point (4) needs
some more explanation: Congestion is defined based on density.
Notwithstanding the difficulties when measuring density, we suggest
density as the most suitable criterion for the identification of
congestion. In addition to the mere occurence of densities exceeding a
certain threshold (say 3.5 persons per square meter), the time this
threshold is exceeded is another necessary condition for a sensible
definition of congestion. In the case presented here, 10\% of the
overall evacuation time is used. Both criteria are in accordance with
the IMO regulations \cite{MSC-Circ.10332002}.

Based on these results, evacuation time and areas of congestion,
corrective actions can be taken. The most straightforward measure would
be a change of geometry, i.e.\ shorter or wider escape paths (floors,
stairs, doors). This can be directly put into the geometry sub-model,
the simulation be re-run, and the result checked. Secondly, the signage
and therefore the orientation capability could be improved. This is not
as straightforward as geometrical changes. It does depend more heavily
on the model characteristics how these changes influence the evacuation
sequence.

We will not go into these details in the following two sections but
rather show two typical examples for evacuation simulations and the
results obtained. We will also not discuss the results in detail, since
they are of an illustrative nature in the context of this article.
The following examples are based on investigations that have been performed 
using a cellular automaton
model which is described along with the simulation program
in \cite{Kluepfel2001,Meyer-Koenig2001}.

\subsubsection{Simulation Example 1 - Hotel}

The first example we show is a hotel with 8069 persons. In
fig.~\ref{fig:initial-distr} only the ground floor is shown. There are
nine floors altogether. The upper floors influence the ground floor
only via the stair landings and the exits adjacent to them. Most of
the 8069 persons are initially located in the ground floor, since the
theatre and conference area is located there. The upper floors are
mainly covering bedrooms and some small conference areas.

\begin{figure*}[thb]
        \centering
                \includegraphics[width=0.45\textwidth]{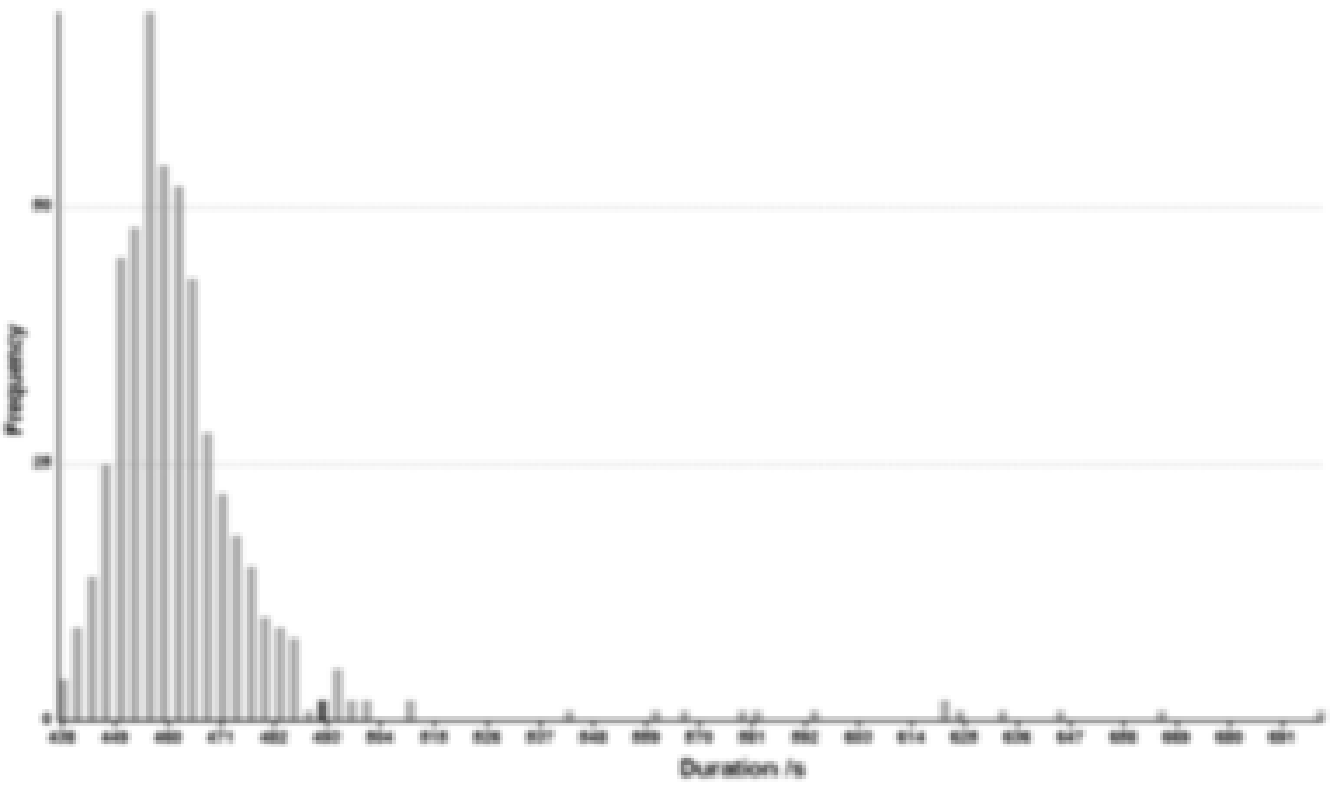}
\qquad
                \includegraphics[width=0.45\textwidth]{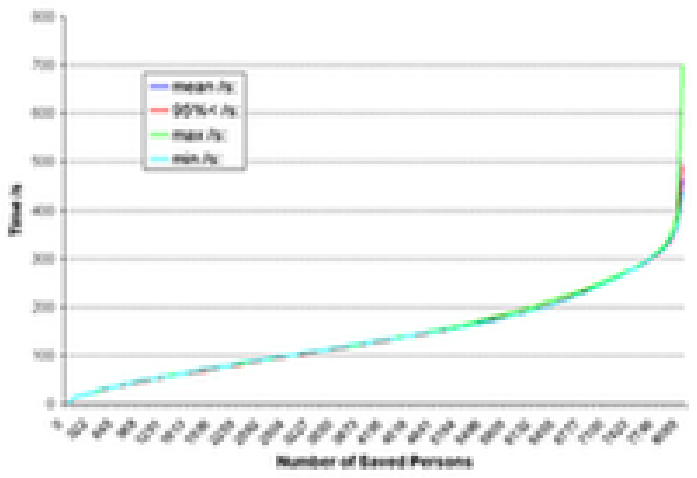}
        \caption{Frequency distribution and evacuation curve.}
        \label{fig:frequency-distribution}
\end{figure*}

\begin{figure*}[thb]
        \centering
                \includegraphics[width=0.45\textwidth]{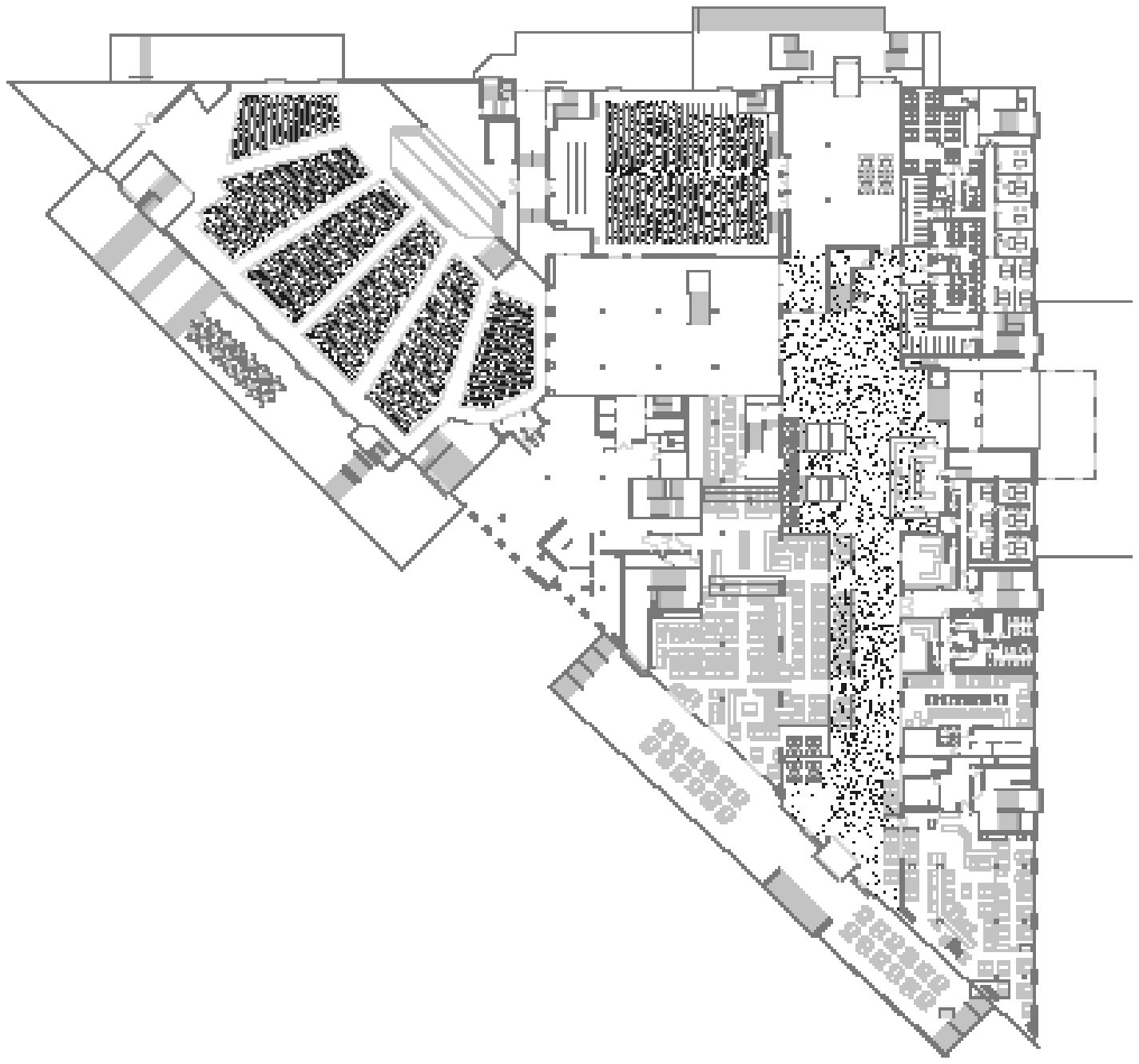}
                \qquad
                \includegraphics[width=0.45\textwidth]{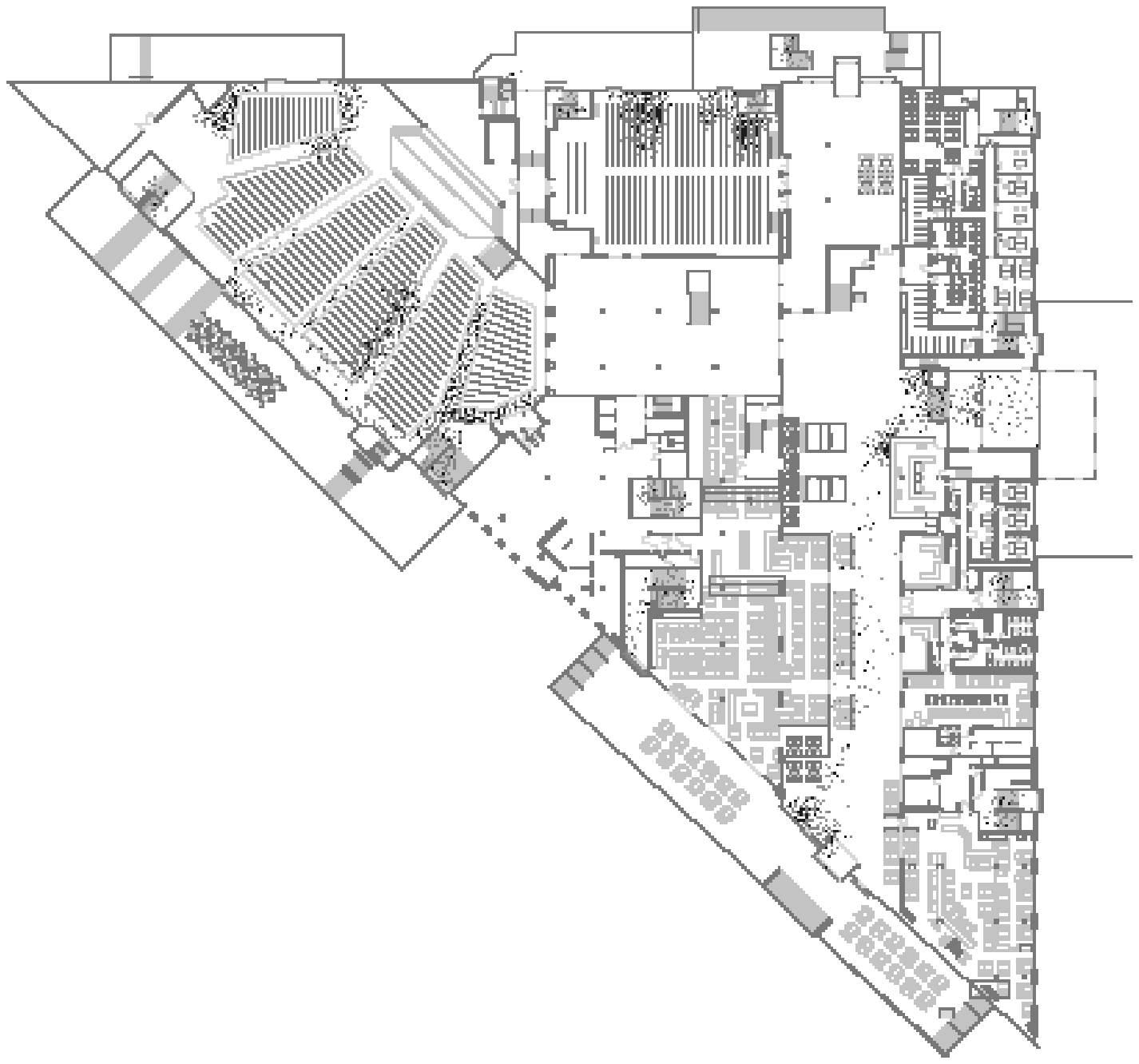}
        \caption{Initial population distribution and situation after 
two minutes.}
        \label{fig:initial-distr}
\end{figure*}

The first step in our example (which might well be a useful recipe for
evacuation analyses in general and is again in accordance with
\cite{MSC-Circ.10332002}) is to perform a statistical analysis. To
this end, 500 samples are simulated. The evacuation time of a
single run is the time it takes for all persons to get out. In this
context, no fire or smoke are taken into account. 
Since there are stochastic influences in the model used, the
significant overall evacuation time is taken to be the 95-percentil
(cf.~fig.~\ref{fig:frequency-distribution}). Finally, the maximum,
minimum, mean, and significant values for the evacuation curve (number
of persons evacuated vs. time) are shown in
fig.~\ref{fig:frequency-distribution}, too.

The next figure (fig.~\ref{fig:density-plot}) shows the cumulated
density. The thresholds (red areas) are 3.5 persons per square meter
and 10\% of the overall evacuation time (in this case 49 seconds). The
overall evacuation time 8:13 minutes (493 seconds). This value is
obtained by taking the 95-percentile of the frequency distribution for
the overall evacuation times
(cf.~fig.~\ref{fig:frequency-distribution}).

\begin{figure}
        \centering
                \includegraphics[width=0.60\textwidth]{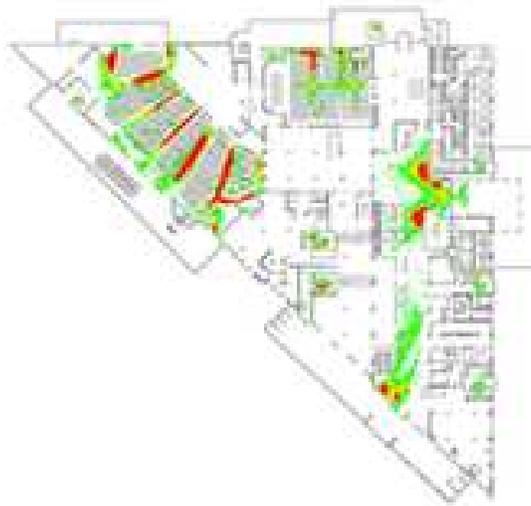}
        \caption{Density plot, i.e. cumulated person density exceeding 
3.5 persons per square meter and 10\% of overall evacuation time.}
        \label{fig:density-plot}
\end{figure}

Of course, a distribution of overall evacuation times (for one scenario,
i.e.\ the same initial parameters) can only be obtained by a stochastic
model. In a deterministic model only one single value is calculated for
the overall evacuation time. The variance of the overall evacuation
times is due to two effects in the model used here: the initial position
of the persons is determined anew at the beginning of each simulation
run  since only the statistical properties of the overall population
is set and the motion of the persons is governed by partially stochastic
rules (e.g.\ probabilistic parameters).

\subsubsection{Simulation Example 2 - Passenger Ship}

The second example we will show is a ship. The major difference to the
previous example is the distinction of (1) assembly phase and (2)
embarkation and launching.
        \[T = A + \frac{2}{3} (E + L) = f_{\rm safety} \cdot (t_{\rm react} 
+ t_{\rm walk}) + \frac{2}{3} (E + L)\ \ \le\ \ 60 \, {\rm minutes}. 
\]
Embarkation and launching time (E+L) are required to be less than 30
minutes. For the sake of the evacuation analysis at an early design
stage, the sum of embarkation and launching time can be assumed to be
30 minutes. Therefore, the requirement for A is 40 minutes.
Alternatively, the embarkation and launching time can be determined by
an evacuation trial.

\begin{figure}
        \centering
        \includegraphics[width=0.31\textwidth]{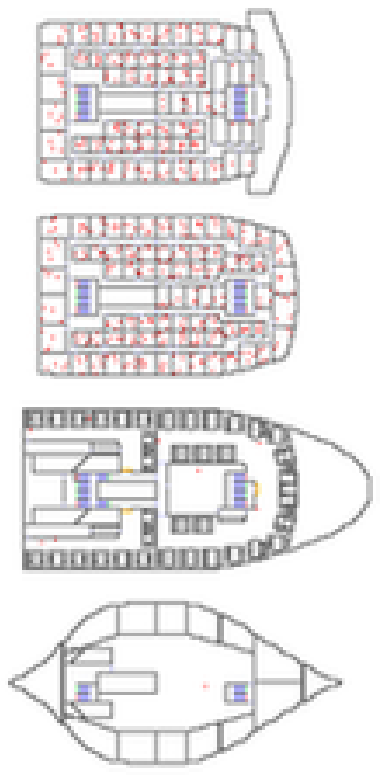}
        \includegraphics[width=0.31\textwidth]{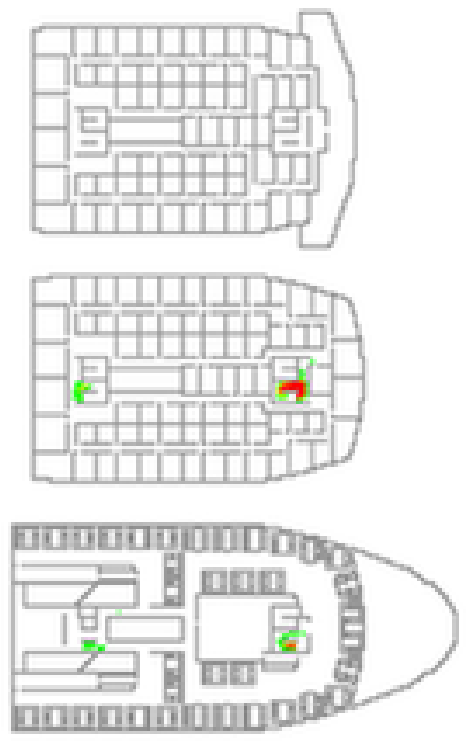}
        \includegraphics[width=0.31\textwidth]{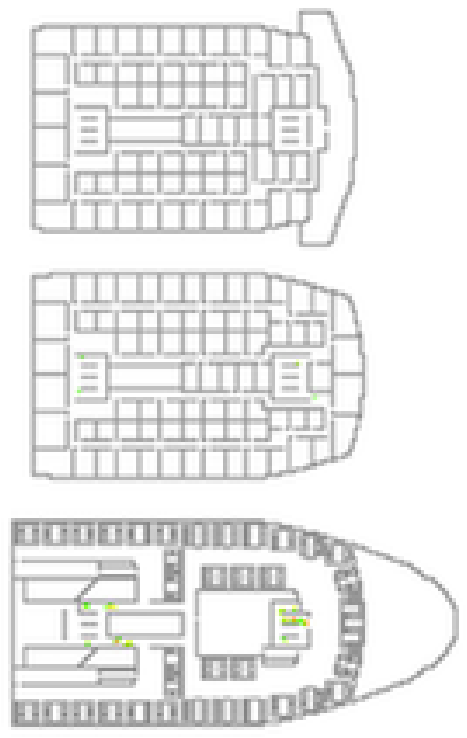}
        \caption{Initial distribution for the night case, density plot 
          for the day case, and density plot for the night case
          for the ``AENEAS steamliner''.}
        \label{fig:aeneas-night-initial}
\end{figure}

Figure~\ref{fig:aeneas-night-initial} shows the layout, initial
population distribution (night case), density plot for the day case,
and density plot for the night case. The reaction times are different
for the day and the night case: 3 to 7 minutes (equally distributed)
in the one and 7 to 13 minutes in the other case. The longer reaction
time in the night case results in less congestion
(cf.~Fig.~\ref{fig:aeneas-night-initial}). 
Both cases have to be done
in the analysis according to \cite{MSC-Circ.10332002}. Additionally, a
secondary night and day case are required (making up for four cases
altogether): In these secondary cases the main vertical zone
(MVZ) leading to the longest overall individual assembly time is
identified, and then either half of the stairway capacity
in this zone is assumed to be not available, or 50\% of the persons
initially located in this zone have to be lead via one neighbouring
zone to the assembly station.

In the same way as shown for the two examples, simulations can be
performed for other types of buildings and vessels. It has been applied
to various passengers ships \cite{als_SCHR02b}
to football stadiums \cite{Kluepfel2006a}
and the World Youth Day 2005 \cite{Kluepfel2006a},
the Jamarat Bridge in Makkah \cite{Kluepfel2006a},
a movie theater and schools (mainly for calibration and validation)
\cite{Kluepfel2001} and airports \cite{SchultzLF06A}.
Of course, many examples of application based on various models can be
found in the literature. For an overview, the proceedings of the PED
conference series are an excellent starting point
\cite{PED01,PED03,PED05}.

\subsection{Comparison of Commercial Software Tools}

From a practical point of view, application of models for pedestrian
dynamics and evacuation processes becomes more and more relevant in
safety analysis. This has lead to the development of a number of
software tools that, with different sophistication, to study many
aspects without risking the health of test persons in evacuation trials.

\hyphenation{Pred-tet-schenski Mi-linski building-EXODUS PedGo}

Commercial and non-commerial software tools are based on different
types of modeling \cite{Kuligowski2005b,Tubbs2007_cr} and they became
very popular since the middle of the 1990s. A first comparison of
different commercial software tools can be found in
\cite{Weckman1999}, where they were attested to produce ``reasonable
results''. Further comparisons of real evacuation data with software
tools or handcalculation methods can be found in
\cite{Ko2003_cr,Ehm2004_cr,Kuligowski2005a,Hoskin2004a,Lord2005_cr,Shestopal1994_cr,Rogsch2007_cr,Rogsch2005_cr}.
But results predicted by different commercial software tools can differ by 
up to 40\% for the same building \cite{Kuligowski2005a}.  
By calculating with different assumptions, e.g.\ different reaction times, 
use of more or less detailed stair models
or calculating with a real occupant load in contrast to an uncertainty 
analysis, the results may be different, too 
\cite{Kuligowski2005a,Lord2005_cr}. Contrary to these results
another study \cite{Rogsch2007_cr} shows that calculations with
different software tools are able to predict total evacuation times
for high-rise buildings and there are no large differences as shown in
\cite{Kuligowski2005a}. In \cite{Rogsch2007_cr} it is also shown that
commercial software tools are not able to predict ``correct''
evacuation times for selected floors of high-rise buildings with very
low densities. 
In this case human behaviour has a very large influence
on the evacuation time contrary to evacuations atwith medium or high
densities, where human behaviour has an smaller influence on the
evacuation time of selected areas, because congestions appear and
continue larger than in low density situations, thus people are
obtaining the exit where the congestion is still alive
\cite{Rogsch2007a_cr}. In low density situations congestions are very
rare, thus people are moving narrowly with free walking velocity
through the building \cite{Rogsch2007a_cr}.
  
\begin{figure}[h]
  \begin{center}
    \includegraphics[width=1.0\textwidth]{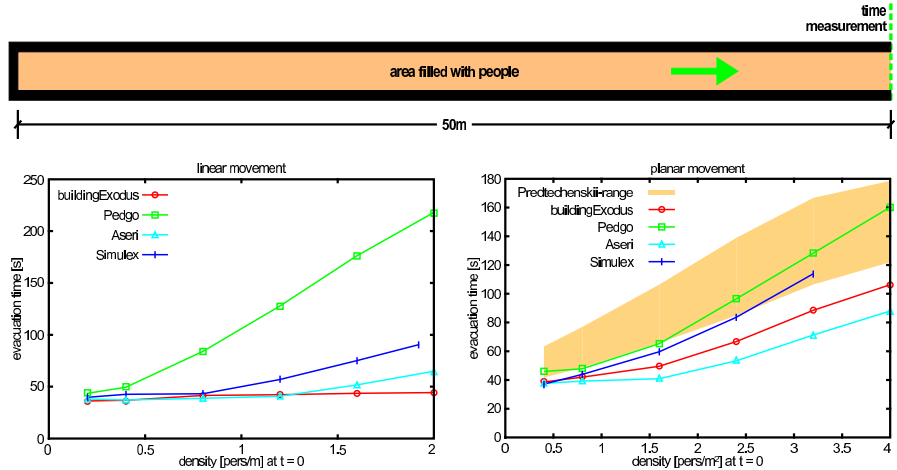}
    \caption{Comparison of different software tools by simulating 
      linear (left) and planar (right) movement \cite{Rogsch2007a_cr}}
\label{fig:floor_cr}
  \end{center}
\end{figure}
\begin{figure}[h]
  \begin{center}
    \includegraphics[width=1.0\textwidth]{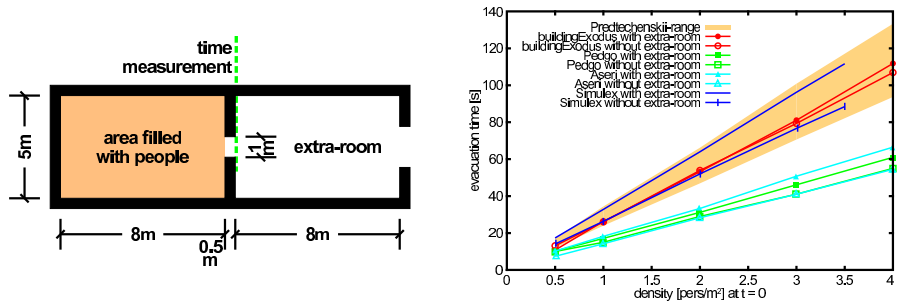}
    \caption{Comparison of different software tools by simulating 
      a simple room geometry \cite{Rogsch2007a_cr}}
\label{fig:simple_room_cr}
  \end{center}
\end{figure}

But the results presented in \cite{Rogsch2007_cr} also show that
commercial software tools have sometimes problems with the empirical
relationship of density and walking speed (see Fig.~\ref{fig:floor_cr}).
Furthermore it is very important how boundary conditions are implemented in
these tools (see Fig.~\ref{fig:simple_room_cr}), and the investigation
of a simple scenario of a single room using different software tools 
shows results differing of about a factor of two 
(see Fig.~\ref{fig:simple_room_cr}) \cite{Rogsch2007_cr}. 
In this case all software tools predict a
congestion at the exit.  Furthermore it is possible that the
implemented algorithm fails \cite{Rogsch2007_cr}. Thus for the user it
is hard to know which algorithms are implemented in closed-source
tools so that such a tool must be considered as ``black box''
\cite{Paulsen1995_cr}. It is also quite difficult to compare results
about density and appearing congestions calculated by different
software tools \cite{Rogsch2007a_cr} and so it is questionable how these
results should be interpreted. But, as pointed out earlier, reliable
empirical data are often missing so that a validation of software tools or
models is quite difficult \cite{Rogsch2007a_cr}.


\section{Future Directions}
\label{sec-futdir}

The discussion has shown that the problem of crowd dynamics and
evacution processes is far from being well understood.
One big problem is still the experimental basis. As in many human
systems it is difficult to perform controlled experiments on a 
sufficiently large scale. This would be necessary since data from
actual emergency situation is usually not available, at least in
sufficient quality.
Progress should be possible by using modern video and computer
technology which should allow in principle to extract precise
data even for the trajectories of individuals.

The full understanding of the complex dynamics of evacuation processes
requires collaboration between engineering, physics, computer science,
psychology etc. Engineering in cooperation with computer science
will lead to an improved empirical basis. Methods from physics allow
to develop simple but realistic models that capture the main aspects
of the dynamics. Psychology is then needed to understand the interactions
between individuals in sufficient detail to get a reliable set of 
`interaction' parameters for the physical models.

In the end these joint efforts will hopefully lead to realistic model
for evacuation processes that not only allow to study these already
in the planning stages of facilities, but even allow for a dynamical
real-time evacuation control in case an emergency occurs.

\subsection*{Acknowledgements}
The authors would like to acknowledge the contribution of Tim Meyer-K\"onig 
(the developer of PedGo) and Michael Schreckenberg, Ansgar Kirchner, 
Bernhard Steffen for many fruitful discussions and valuable hints. 

\phantom{\cite{ACRI06A,PED01,PED03}}

\newpage

\section{Bibliography}

{\large \bf Books and Reviews} \\
\begin{enumerate}
\item
  V.M. Predtechenskii and A.I. Milinskii:
  \emph{Planing for foot traffic flow in buildings},
Amerint Publishing, New Delhi (1978)

\item
P.J. DiNenno (Ed.): 
\emph{SFPE Handbook of Fire Protection Engineering},
National Fire Protection Association (2002)

\item
M. Schreckenberg and S.D. Sharma (Eds.):
\emph{Pedestrian and Evacuation Dynamics}
Springer (2002)

\item
E.R. Galea (Ed.):
\emph{Pedestrian and Evacuation Dynamics '03}
CMS Press, London (2003)

\item
N. Waldau, P. Gattermann, H. Knoflacher, and M. Schreckenberg  (Eds.):
\emph{Pedestrian and Evacuation Dynamics '05}
Springer (2007)

\item 
A. Schadschneider, T. P\"oschel, R. K\"uhne, M. Schreckenberg, and
D.E. Wolf (Eds.): \emph{Traffic and Granular Flow '05}
Springer (2007) (see also previous issues of this conference series)
\item
J.S. Tubbs and B.J. Meacham:
\emph{Egress Design Solution - A Guide to Evacuation and Crowd 
Management Planning},
Wiley and Sons (2007)

\item 
D. Chowdhury, K. Nishinari, L. Santen, and A. Schadschneider:
\emph{Stochastic transport in complex systems: From molecules to vehicles},
Elsevier (2008)

\item
Webpage of {\tt Ped-Net} collaboration:  
{\tt www.ped-net.org}  (including discussion forum)

\item
   B. Chopard and M. Droz:
   \emph{Cellular automaton modeling of physical systems},
   Cambridge University Press, Cambridge  (1998)

\end{enumerate}
\vspace{1cm}

\renewcommand{\refname}{Primary Literature}

\bibliographystyle{plain}

\renewcommand{\bibname}{Primary Literature}



\end{document}